\documentclass[useAMS,usenatbib]{mn2e}

\usepackage[flushleft]{threeparttable}
\usepackage{graphicx}
\usepackage{xtab,afterpage}
\usepackage{amsmath}
\usepackage{epstopdf}
\usepackage{url}
\usepackage[flushleft]{threeparttable}
\usepackage{booktabs,fixltx2e}
\usepackage{amssymb}

\newcommand{\icarus}{Icarus}

\newcommand{\pasp}{{\it PASP~\/}}

\newcommand{\apj}{ApJ}
\newcommand{\apjl}{ApJ}

\newcommand{\aap}{A \& A}
\newcommand{\araa}{ARA\&A}
\newcommand{\aj}{AJ}
\newcommand{\mnras}{MNRAS}

\newcommand{\nat}{Nature}

\def\jgr{\rmfamily{J.~Geophys.~Res.~}}

\renewcommand{\v}{\ensuremath{\mathbf{v}}}

\makeatletter
\newcommand*{\rom}[1]{\expandafter\@slowromancap\romannumeral #1@}
\makeatother

\title[Hybrid discs could be of secondary origin]{Imaging [CI] around HD 131835: reinterpreting young debris discs with protoplanetary disc levels of CO gas as shielded secondary discs}
\author[Q. Kral et al.]{Quentin Kral,$^{1,2}$\thanks{E-mail: qkral@ast.cam.ac.uk} Sebastian Marino, $^{1}$ Mark C. Wyatt,$^{1}$ Mihkel Kama,$^{1}$ and
\newauthor{Luca Matr{\`a},$^{3}$}\\
$^{1}$Institute of Astronomy, University of Cambridge, Madingley Road, Cambridge CB3 0HA, UK\\
$^{2}$LESIA, Observatoire de Paris, Universit{\'e} PSL, CNRS, Sorbonne Universit{\'e}, Univ. Paris Diderot,\\ Sorbonne Paris Cit{\'e}, 5 place Jules Janssen, 92195 Meudon, France\\
$^{3}$Harvard-Smithsonian Center for Astrophysics, 60 Garden Street, Cambridge, MA 02138, USA\\
}

\begin{document}

\date{Accepted 1928 December 15. Received 1928 December 14; in original form 1928 October 11}

\pagerange{\pageref{firstpage}--\pageref{lastpage}} \pubyear{2002}

\maketitle

\label{firstpage}

\begin{abstract}
Despite being $>10$Myr, there are $\sim$10 debris discs with as much CO gas as in protoplanetary discs. Such discs have been assumed to be ``hybrid'', i.e., with secondary dust but primordial gas. Here we show that both the dust and gas in such systems could instead be secondary, with the high CO content caused by accumulation of neutral carbon (C$^0$) that shields CO from photodissociating; i.e., these could be ``shielded secondary discs''. New ALMA observations are presented of HD131835 that detect $\sim 3 \times 10^{-3}$ M$_\oplus$ of C$^0$, the majority 40-200au from the star, in sufficient quantity to shield the previously detected CO. A simple semi-analytic model for the evolution of CO, C and O originating in a volatile-rich planetesimal belt shows how CO shielding becomes important when the viscous evolution is slow (low $\alpha$ parameter) and/or the CO production rate is high. Shielding by C$^0$ may also cause the CO content to reach levels at which CO self-shields, and the gas disc may become massive enough to affect the dust evolution. Application to the HD 131835 observations shows these can be explained if $\alpha \sim 10^{-3}$; an inner cavity in C$^0$ and CO may also mean the system has yet to reach steady state. Application to other debris discs with high CO content finds general agreement for $\alpha=10^{-3}$ to $0.1$. The shielded secondary nature of these gas discs can be tested by searching for C$^0$, as well as CN, N$_2$ and CH$^{+}$, which are also expected to be shielded by C$^0$.
\end{abstract}


\section{Introduction}

Gas is now detected routinely around main sequence stars that possess planetesimal belts similar to the Kuiper belt. On the order of 20 systems show the presence of gas \citep[e.g.,][]{2013ApJ...776...77K,2016ApJ...828...25L,2017ApJ...839...86H,2016MNRAS.461.3910G,2017ApJ...849..123M} and this number will soon increase, mostly thanks to upcoming surveys with ALMA. Most of these systems have CO detected
in emission (often colocated with the planetesimal belt) and for a handful of them ionised carbon and oxygen were detected with Herschel \citep[e.g.,][]{2012A&A...546L...8R,2014A&A...565A..68R,2016A&A...591A..27B,2017haex.bookE.165K}. Neutral carbon (C$^0$) has been predicted to be a good tracer of gas in these systems \citep{2017MNRAS.469..521K} and has now been detected around 49 Ceti and $\beta$ Pic \citep{2017ApJ...839L..14H, Catal}.

The origin of the gas around these main sequence stars is still debated for some systems (because of their youth that can be close to 10 Myr or older), i.e., whether the gas is a remnant of the protoplanetary disc (called primordial hereafter) or released at a later stage (i.e., secondary). However, for at least three systems (HD 181327, $\beta$ Pic and Fomalhaut) the CO mass detected is so low that it cannot be primordial as neither CO 
nor H$_2$ (even assuming an extreme CO-to-H$_2$ ratio of $10^{-6}$) could shield CO from photodissociating owing to the UV radiation from the interstellar radiation field \citep{2016MNRAS.460.2933M,2017MNRAS.464.1415M,2017ApJ...842....9M}.  
We also note that the presence of gas around the 440 Myr old Fomalhaut star \citep{2017ApJ...842....9M} and possibly around the 1 Gyr old $\eta$ Corvi \citep{2017MNRAS.465.2595M} leaves no doubt concerning the secondary origin of the gas there. 

\citet{2017MNRAS.469..521K}
show that a second generation model where the gas is released from volatile-rich planetesimals colliding in the belts \citep[e.g.][]{2012ApJ...758...77Z} can explain most of the detections and non-detections so far (for $>$10 Myr systems), assuming planetesimals with a composition similar to Solar System comets, 
reinforcing the idea that most of the observed gas is secondary. However, the high-mass gas discs, called hybrid 
discs by \citet{2013ApJ...776...77K} because of their high, assumed primordial, CO content but low debris-disc like secondary dust mass, could not be explained with this
model and were deemed to be of primordial origin. 

We consider in this paper whether these high gas mass systems can actually also be explained as being of secondary origin due to an ingredient that was missing in the previous model: {\it the C$^0$ shielding effect}. \footnote{We note that
C$^0$ shielding was taken into account in the numerical model by \citet{2016MNRAS.461..845K} studying $\beta$ Pic, but only for its impact on ionisation fraction and temperature, but not for the CO content as the C$^0$ mass for $\beta$ Pic is not large enough to shield CO. \citet{2017MNRAS.464.1415M}
showed later that indeed in $\beta$ Pic, the shielded CO photodissociation rate due to C$^0$ is 0.96 of the unshielded one, i.e. not important.}
This effect is now accounted for in a new model presented in this paper and applied to a
new [CI] ALMA detection towards HD 131835 also presented in this paper.

HD 131835 is a 15-16 Myr old \citep{2002AJ....124.1670M,2012ApJ...746..154P} A-type star located at a distance\footnote{We note that the Hipparcos distance assumed in previous studies was $122.75^{+16.2}_{-12.8}$, which is consistent
within error bars with the GAIA DR1 value, but the median ratio is 1.18 \citep{2007A&A...474..653V}, which can affect some earlier derived model values such as dust or gas masses, stellar luminosities, ... In particular, distances from previous studies should be multiplied by 1.18 before comparing to our study.} of 145$\pm$9 pc \citep{2016A&A...595A...1G,2016A&A...595A...2G} that is probably a member of the Upper Centaurus Lupus moving group \citep[a subgroup of the Sco-Cen association,][]{2011MNRAS.416.3108R}. An infrared (IR) excess
around this star has first been reported by \citet{2006ApJ...644..525M}, establishing the presence of a bright debris disc. The disc was first resolved in mid-infrared using T-ReCS from which it could be deduced that multiple disc components were needed to fit the data 
\citep{2015ApJ...815L..14H}. More recently, SPHERE observations in the near-IR led to the detection of a series of narrow concentric rings \citep{2017A&A...601A...7F}. Such concentric rings may be explained by the presence of large quantities of gas \citep[e.g.][]{2001ApJ...557..990T,2013Natur.499..184L,2018ApJ...856...41R}.

CO gas has recently been discovered around HD 131835 using APEX \citep{2015ApJ...814...42M} and has been followed-up with ALMA from which the total CO mass derived (using an optically thin line) is $\sim$ 0.04 M$_\oplus$ \citep[after correction for the new GAIA distance,][]{2017ApJ...849..123M}. This CO mass is
of the same order of magnitude as for the famous TW Hya protoplanetary disc \citep{2013ApJ...776L..38F}, leading to the possibility that the disc is primordial.

From the new [CI] observations we present in this paper, we show that the observed gas in the so-called ``hybrid'' disc around HD 131835 (or HIP 73145) is compatible with being a shielded debris disc, i.e., a massive gas disc of secondary origin.
Furthermore, we conclude that some or all of the so-called hybrid discs might not be hybrid after all but just entirely second generation, both in their gas and dust content.
For this reason, from now on we will call these discs: {\it shielded discs} rather than hybrid. To make the distinction easier, we call the low CO mass systems of secondary origin ``unshielded'' and the high CO mass systems ``shielded''.
This has important consequences for our understanding of the fate of protoplanetary discs that, if true, would not survive as long as previously thought and would indicate that massive secondary gaseous debris discs can start early, while planet formation may still be on-going.

In Sec.~\ref{shield}, we start by introducing the new concept of shielded discs that can explain massive CO gas discs as being of secondary origin rather than hybrid.
In Sec.~\ref{obs}, we present new continuum observations of HD 131835 with ALMA and a fit of the data with a dust model. We then carry on by showing the 
first neutral carbon detections of HD 131835 with ALMA and a fit of the data set with several gas models in Sec.~\ref{gas}. In section \ref{discu}, we discuss the implications of this new [CI] detection and how it backs up the shielded disc scenario. 
We first describe a self-consistent physical model that can explain the observed high CO mass together with the new [CI] detection assuming a secondary origin of the gas. We then constrain the viscosity of the observed gas disc (using an $\alpha$ parametrization), derive the expected composition of the (exo)planetesimals in this system and discuss the dust/gas
interactions that may happen. Finally, before concluding, we discuss the possibility that all hybrid discs discovered so far (possibly up to 9 systems) are in fact shielded discs, i.e., not hybrid, and we derive some constraints on their gas disc viscosity ($\alpha$) and predict their content 
in C$^0$ that is within reach of ALMA.
 
\section{Secondary gas discs and introduction to the new category of shielded discs}\label{shield}

\subsection{Standard evolution of low-mass unshielded secondary gas discs}
\begin{figure}
   \centering
   \includegraphics[width=8cm]{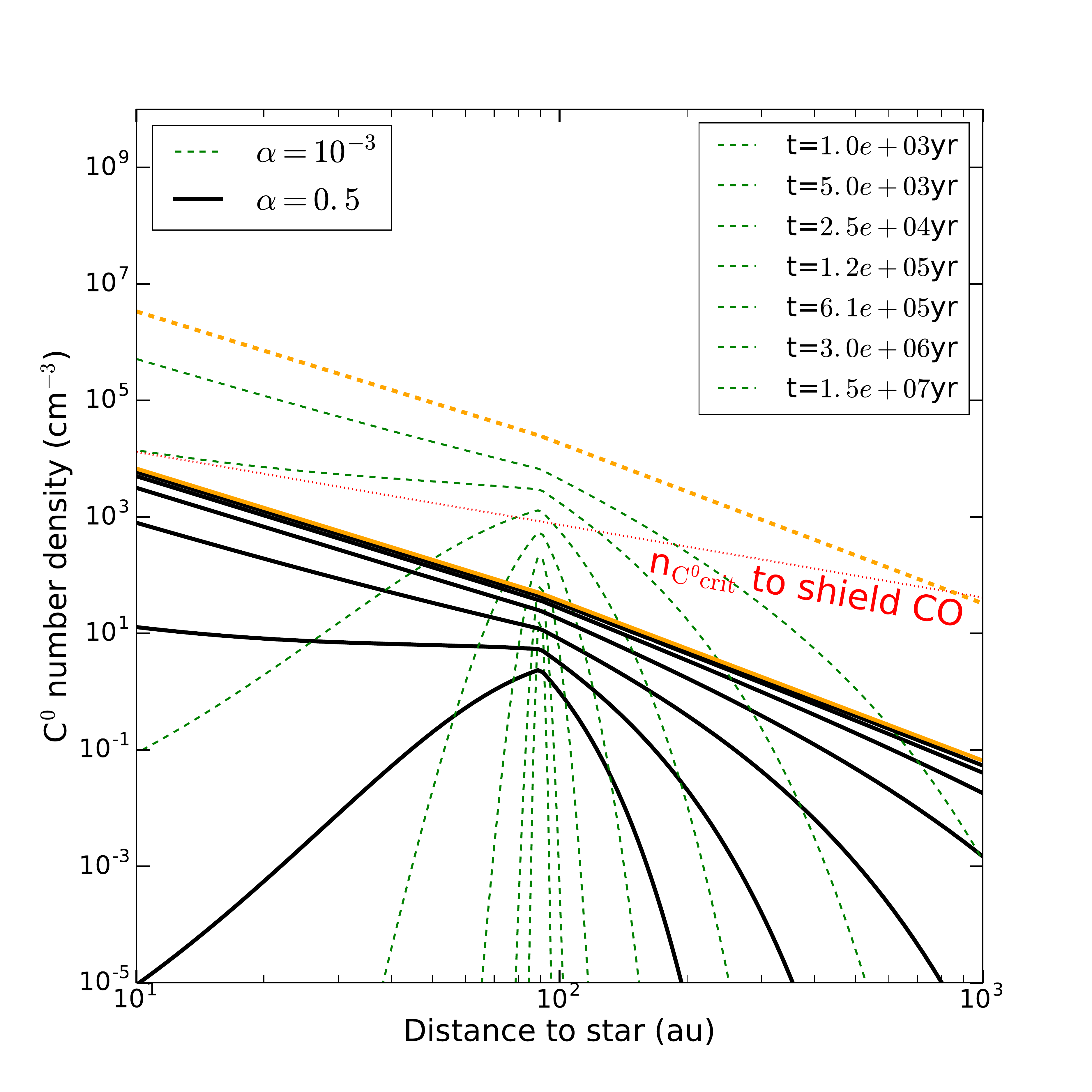}
   \caption{\label{figevol} Evolution of the C$^0$ number density as a function of radius and for several timesteps (from $10^3$ yr to 15 Myr) for carbon inputted at a steady rate at 90 au. The carbon gas is assumed to spread viscously with time with a timescale that depends on the viscosity parametrized 
by $\alpha$. We show a fast evolution with $\alpha=0.5$ (black solid lines) and a slower evolution with $\alpha=10^{-3}$ (green dashed lines). The high-$\alpha$ disc reaches a steady state after a few Myr whilst the low-$\alpha$ disc is still evolving after 15 Myr. The red dotted line shows
the critical C$^0$ density above which C$^0$ starts shielding CO from photodissociation. For the low-$\alpha$ case, we note that the C$^0$ density is high enough to extend the CO lifetime and CO will accumulate owing to C$^0$ shielding, which will lead to a shielded disc as defined in Sec.~\ref{shield}. 
To compute the C$^0$ input rate, we assumed $\dot{M}_{\rm CO}=10^{-2}$ M$_\oplus$ and an ionisation fraction of 0.1. The orange lines show the steady state number density for $\alpha=0.5$ (solid) and $10^{-3}$ (dashed) based on Eq.~\ref{Sig}.}
\end{figure}

In a secondary gas production scenario, gas is released in the planetesimal belt, while volatile-rich bodies collide with each other. CO and water are expected to be the major constituents of comets (assuming Solar System-like compositions), and will possibly dominate the release \citep[note that this depends on the exact mechanism that releases this gas, see][]{2018ApJ...853..147M}. In the standard secondary approach 
presented in \citet{2016MNRAS.461..845K,2017MNRAS.469..521K}, CO and water photodissociate very quickly because of the UV radiation from the interstellar radiation field (i.e., typically in about 100 yr) because there is not enough CO to self-shield 
nor enough H$_2$ that could increase the CO or water lifetime. Carbon, oxygen and hydrogen are thus created very quickly through photodissociation, which builds an atomic gas disc that viscously spreads. 
In our model, the viscosity is parametrized with an $\alpha$ parameter \citep[as in][]{2013ApJ...762..114X}, which has been inferred to be high in the $\beta$ Pic gas disc, i.e., $\alpha>0.1$, leading to a viscous timescale $t_\nu < 0.5$ Myr \citep{2016MNRAS.461..845K} as 
the viscous evolution timescale at $R_0$, $t_\nu(R_0)$ is given by $R_0^2 \Omega/(\alpha c_s^2)$, where the orbital frequency $\Omega$ and the sound speed $c_s$ are both estimated at $R_0$ (taken to be 85 au for $\beta$ Pic). However, in systems in which the carbon ionisation fraction is smaller, and therefore the magnetorotational instability less effective \citep{2016MNRAS.461.1614K}, the resulting $\alpha$ could be much smaller (e.g., $10^{-3}$), resulting in a much longer viscous timescale.

To illustrate the implication of a change in $\alpha$ for the evolution of the disc, Fig.~\ref{figevol} shows the temporal evolution of a gas disc with a steady input of gas at a rate of 
$\dot{M}_{\rm CO}=10^{-2}$ M$_\oplus$/Myr at $R_0=90$ au, which is converted to a C$^0$ input rate by assuming an ionisation fraction $f$ of 0.1. The evolution is worked out semi-analytically using solutions found by \citet{2011MNRAS.410.1007T}, where we assumed the temperature
to be $\propto R^{-\beta}$, with $\beta=1/2$.
For a $\beta$ Pic-like system with $\alpha=0.5$ (solid black lines), the steady state is reached after 
only a few Myr (i.e., longer timescale lines all coincide on the steady state uppermost solid black line). For a model in which $\alpha=10^{-3}$ (green dashed lines), the steady state is not yet reached after 15 Myr (last epoch shown in the figure). In this case, the evolution is 500
times slower and much more carbon can accumulate in the system, reaching number densities $n_{{\rm C}^0}$ of $\sim 10^4$ cm$^{-3}$ at $R_0$ after 15 Myr. 
We also plot the predicted analytical steady state values that we find to be \citep[updating the solutions for the case $\beta=1/2$ by][to our case of a steady input]{1974MNRAS.168..603L}

\begin{equation}\label{Sig}
   \Sigma(R) = \frac{\dot{M}}{3 \pi \nu_0}
    \begin{cases}
      \left(\frac{R}{R_0}\right)^{\beta-3/2} & \text{for $R<R_0$}\\
      \left(\frac{R}{R_0}\right)^{\beta-2} & \text{for $R>R_0$}\\
    \end{cases}       ,
\end{equation}

\noindent where $\nu_0=\alpha c_s^2(R_0)/\Omega(R_0)$ is the viscosity at $R_0$, the temperature $T \propto R^{-\beta}$, and $\dot{M}$ is the input rate (for C$^0$, $\dot{M}=12/28 \dot{M}_{\rm CO} (1-f)$). \citet{2012MNRAS.423..505M} also derive this 
formula (in their appendices) and show that it is valid not only for $\beta=1/2$ but for all values of $\beta$. Therefore, at steady state the number density scales as $R^{3\beta/2  -3}$ for $R<R_0$ and as $R^{3\beta/2 -7/2}$ for $R>R_0$ as shown by the orange analytical steady state profiles
plotted in Fig.~\ref{figevol}.

\subsection{Shielded discs}\label{shieldeddd}

\begin{figure}
   \centering
   \includegraphics[width=8cm]{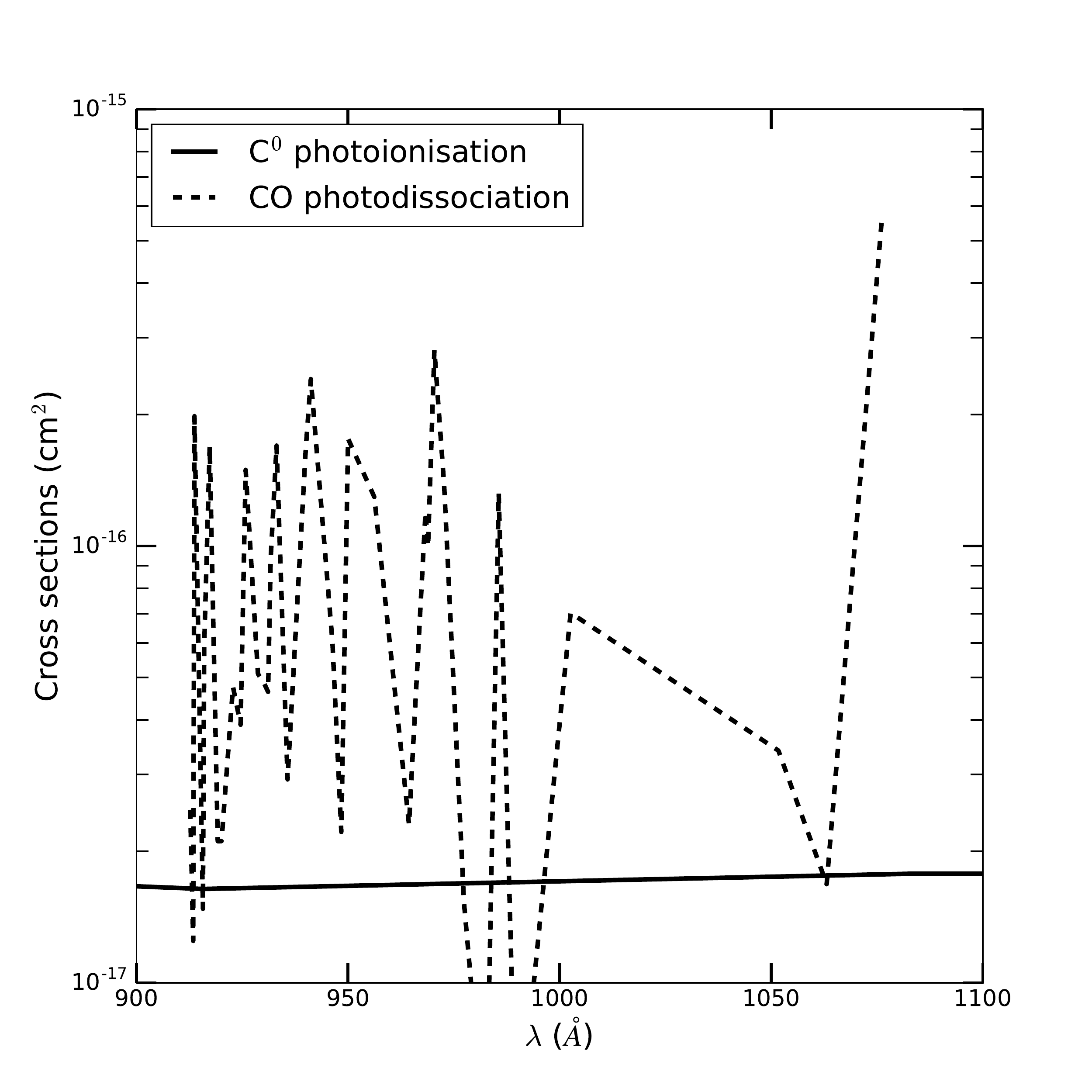}
   \caption{\label{figphd} C$^0$ photoionisation \citep[solid,][]{1988ASSL..146...49V} and CO photodissociation \citep[dashed,][]{2009A&A...503..323V} cross sections between 900 and 1100~\AA. The same energetic photons that can photodissociate CO can photoionise neutral carbon.}
\end{figure}

The shielded disc case happens when the C$^0$ density becomes high enough that C$^0$ gets optically thick to UV radiation. The C$^0$ photoionisation cross section is roughly constant (see Fig.~\ref{figphd}) and equal to $\sigma_i=1.6\times10^{-17}$ cm$^2$ \citep{1988ASSL..146...49V} from
900$\,$\AA$\,$  (i.e., close to the Lyman break at 13.6eV) to $\sim 1100\,$\AA$\,$ (i.e., 11.26eV, which is the ionisation potential of carbon). 
Fig.~\ref{figphd} shows that the photodissociation cross sections of CO \citep{2009A&A...503..323V} are affected by the same photons (i.e., from 910 to 1075$\,$\AA) that can ionise carbon. 
Calculating the C$^0$ column density $N_{{\rm C}^0}$ from $n_{{\rm C}^0}$ in Fig.~\ref{figevol} for the $\alpha=10^{-3}$ model at 15 Myr, 
we find that for a UV photon from the interstellar radiation field \citep[IRF, that dominates the UV flux at the typical distances of typical planetesimal belts $R_0$,][]{2018ApJ...859...72M} travelling in the vertical direction,
C$^0$ is indeed optically thick to UV radiation (we note that it is much more optically thick in the radial direction and the star UV field is thus not expected to contribute to CO photodissociation in these shielded discs). 
This means that if C$^0$ is blocking the UV photons, CO will photodissociate on longer timescales.
Indeed, \citet{2012MNRAS.427.2328R} compute that the CO photodissociation timescale (in yr) in the presence of C$^0$ is

\begin{equation}\label{tco}
 t_{\rm CO}=120 \, {\rm yr} \times \exp(\sigma_i N_{{\rm C}^0})
\end{equation}

\noindent where 120 yr is the photodissociation timescale without shielding, i.e., when only the IRF 
is taken into account. This shows that shielding of CO by C$^0$ starts happening for $N_{{\rm C}^0} \gtrsim 10^{16}$ cm$^{-2}$. We note that we considered shielding in the vertical direction and shielding in the radial direction is expected to be much higher because of the greater column density towards the star.

\begin{figure}
   \centering
   \includegraphics[width=8cm]{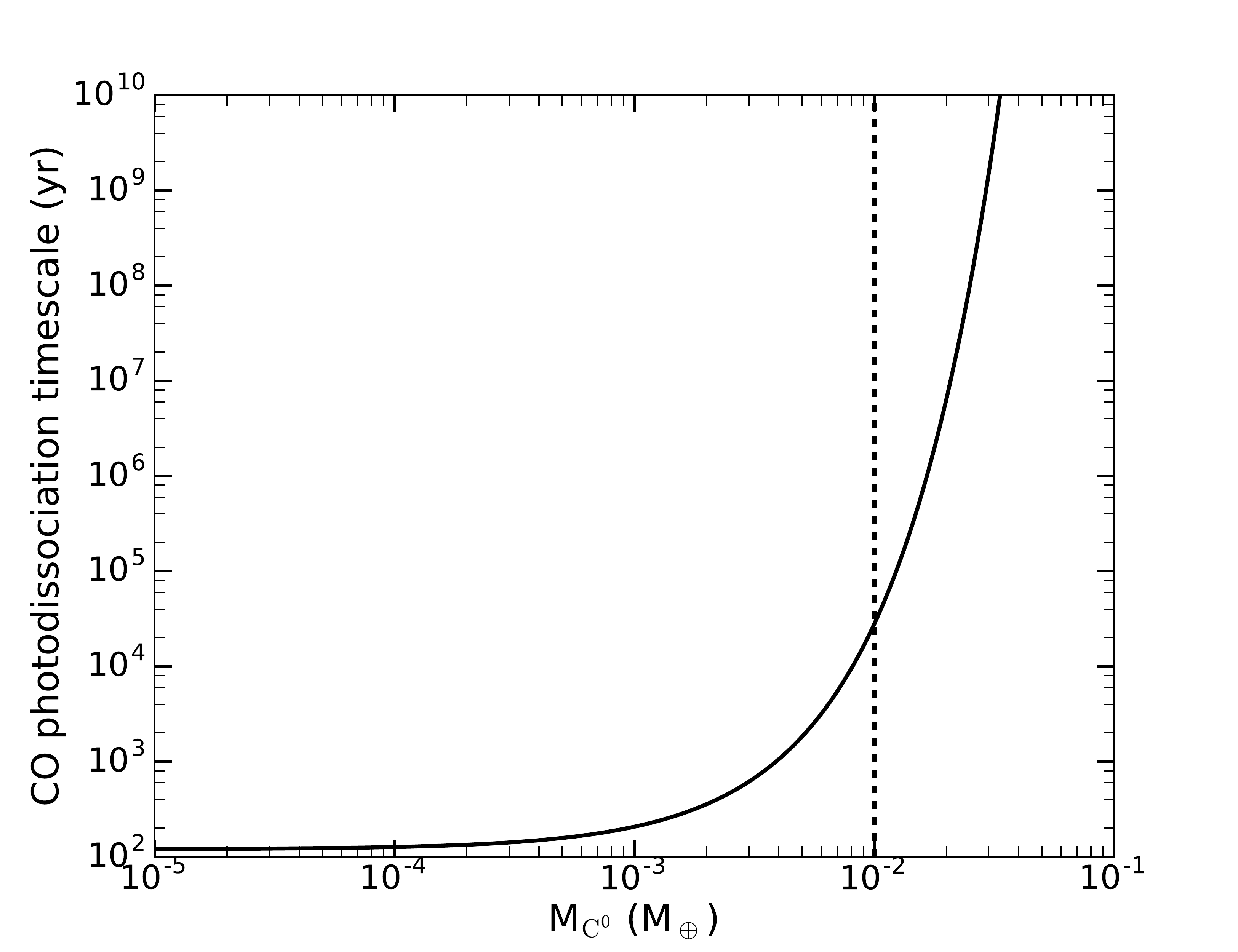}
   \caption{\label{figCIneed} CO photodissociation timescale (in yr) as a function of the C$^0$ mass (M$_{{\rm C}^0}$ in M$_\oplus$) in the gas disc. This timescale depends exponentially on C$^0$ mass and a small amount of C$^0$ can create a large accumulation of CO that is then very hard to photodissociate because the UV photons
that could potentially do that are mostly absorbed by C$^0$ beforehand (see also Fig.~\ref{figphd}). We assumed a disc located at 90 au of width 70 au. The dashed line is for a C$^0$ mass of $10^{-2}$ M$_\oplus$ for which the photodissociation timescale is already more than 100 times
longer than when just assuming the interstellar radiation field ($\sim$120 yr).}
\end{figure}

Fig.~\ref{figCIneed} shows how much C$^0$ mass is needed for the CO 
photodissociation timescale to depart from 120 yr. Here, we also assume that $R_0=90$ au and a width of 70 au so that $M_{{\rm C}^0}=N_{{\rm C}^0} 2 \pi R \Delta R \mu_c m_H$, where $\mu_c$ is the mean molecular weight of carbon.
For a C$^0$ mass of $\sim 10^{-3}$ M$_\oplus$, $t_{\rm CO}$ is 200 yr and as the evolution follows a very steep exponential profile, for $M_{{\rm C}^0}=1.2 \times 10^{-2}$ M$_\oplus$, then $t_{\rm CO}=10^5$ yr.
For the latter case, this means that the CO mass expected from second generation models such as described in \citet{2017MNRAS.469..521K} can be more than $t_{\rm CO}/120 \sim 10^3$ times larger than previously thought 
and one can accumulate CO masses of the order of the mass content found in a protoplanetary disc such as TW Hydra \citep{2013ApJ...776L..38F}.

\begin{figure}
   \centering
   \includegraphics[width=8cm]{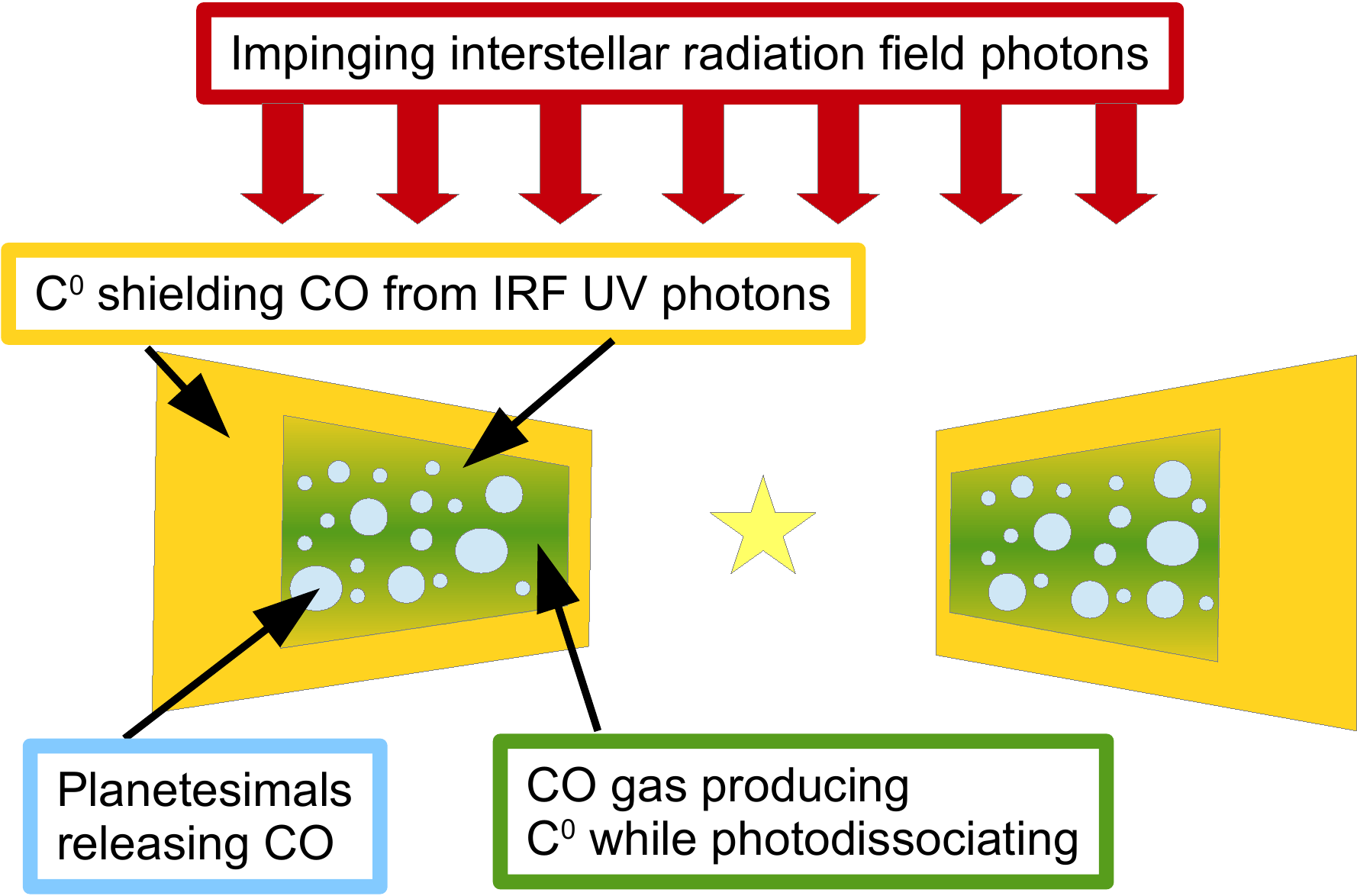}
   \caption{\label{figcartoon} Sketch of the secondary gas model, where 1) gas is released from the solid bodies in a Kuiper-belt like disc, 2) C$^0$ is produced by photodissociation of CO and viscously spread towards the inner region, 3) the C$^0$ gas disc can become massive
enough that it absorbs most photons from the interstellar radiation field that could have photodissociated CO so that CO accumulates and can reach protoplanetary CO mass level. Note that carbon is also present in the green region and that it is more extended outwards than CO.}
\end{figure}

Therefore, the high CO mass content ($\sim0.04$ M$_\oplus$) found by \citet{2017ApJ...849..123M} for HD 131835 could be compatible with a gas production of secondary origin for high enough C$^0$ masses in the disc as summarised by the sketch presented in Fig.~\ref{figcartoon}.
This will be checked in more detail in Sec.~\ref{discu} presenting some new modelling of this C$^0$ shielding effect but first let us check the spatial distribution of the dust (Sec.~\ref{obs}) and gas from new ALMA observations and whether there is enough C$^0$ observed to produce this shielding (Sec.~\ref{gas}).

\section{ALMA continuum observation of HD 131835}\label{obs}
HD 131835 was observed by ALMA in band 8 on 22$^{\rm nd}$ March 2017 as part of the cycle 4 project 2016.1.0.01253. The observations were carried out using 42 antennas with baselines ranging from 15 to 160m.
The total observing time on source (excluding overheads) was 14.1min and the mean PWV was 0.57mm.

Out of the four spectral windows provided by the ALMA correlator, three focused on observing the dust continuum with 128 channels centred at 480.201, 482.159 and 494.201 GHz (bandwidth of 2 GHz). 
The fourth spectral window targeted the CI ($^{3}$P$_0$-$^{3}$P$_1$) transition at 492.160651 GHz (i.e., a rest wavelength of 609.135 $\mu$m) with a higher spectral resolution of 488.281 kHz (0.297 km/s at the rest frenquency of the line) over 3840 channels (i.e., a bandwidth of 1.875 GHz) centred at 492.201 GHz.
Titan was used as a flux calibrator, while J1427-4206 and J1454-3747 were used as bandpass and phase calibrator, respectively.
We used the pipeline provided by ALMA to apply calibrations.

\subsection{Dust continuum observations}
\begin{figure}
   \centering
   \includegraphics[width=8cm]{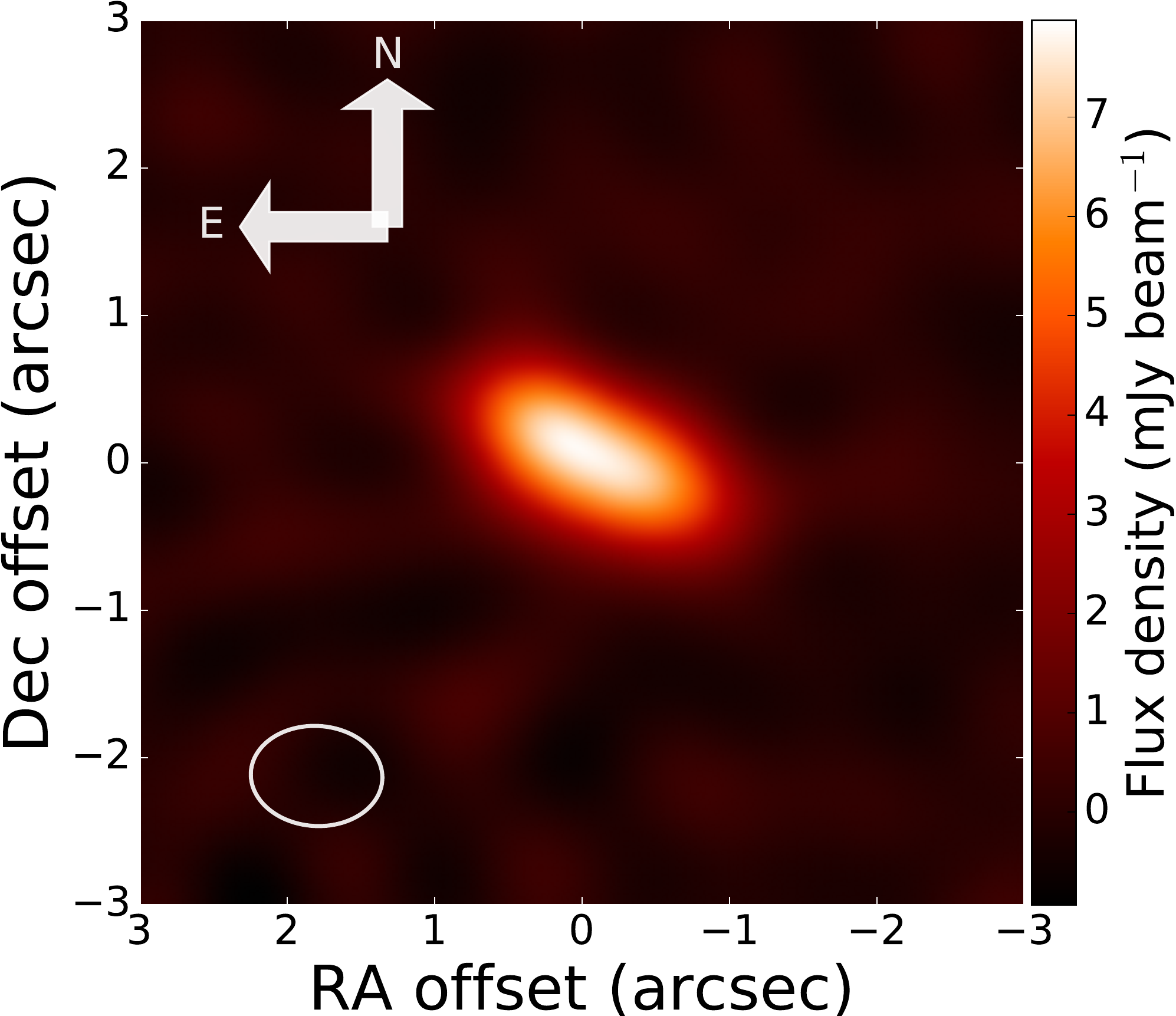}
   \caption{\label{figcont} ALMA dust continuum map of HD 131835 at 487 GHz (band 8). The image is Briggs-weighted (with robust=0.5). The ellipse shows the beam of 0.89 $\times$ 0.68 arcsec and the North and East directions are indicated by the two arrows.
The x- and y-axes indicate the offset from the stellar position in RA and Dec (in arcsec).}
\end{figure}

Fig.~\ref{figcont} shows the ALMA dust continuum map of HD 131835 at $\sim$487 GHz (i.e., 616 $\mu$m). All the analysis that follows uses the visibilities but we show the image to introduce the data to the reader and provide some first qualitative idea of the observation. We
deconvolved the image using the CLEAN algorithm and Briggs weighting with a robust parameter of 0.5. This yields a synthesised beam of  0.89 $\times$ 0.68 arcsec (i.e., $\sim 129 \times 97$ au, assuming a distance of 145 pc from Earth). The peak signal-to-noise ratio (S/N)
in the image is $\sim$32. The image rms around the disc is $\sigma=250\,\mu$Jy/beam. The total flux in the Briggs-weighted image is 17.9$\pm$2.1 mJy, where the error comes from the image noise and flux calibration uncertainties ($\sim$ 10\%) added in quadrature.

The bulk of the emission is within 1 arcsec (i.e., $\sim$145 au) and appears consistent with a highly inclined axisymmetric disc as found from previous resolved observations of the disc at different wavelengths \citep[e.g.][]{2015ApJ...815L..14H,2017A&A...601A...7F}.
The disc parameters such as inclination, PA, and disc mass are derived in the next subsection~\ref{dustfit} using an MCMC approach to fit a dust disc model to the observed visibilities.

\subsection{Fit of the continuum data}\label{dustfit}

\begin{figure*}
   \centering
   \includegraphics[width=15cm]{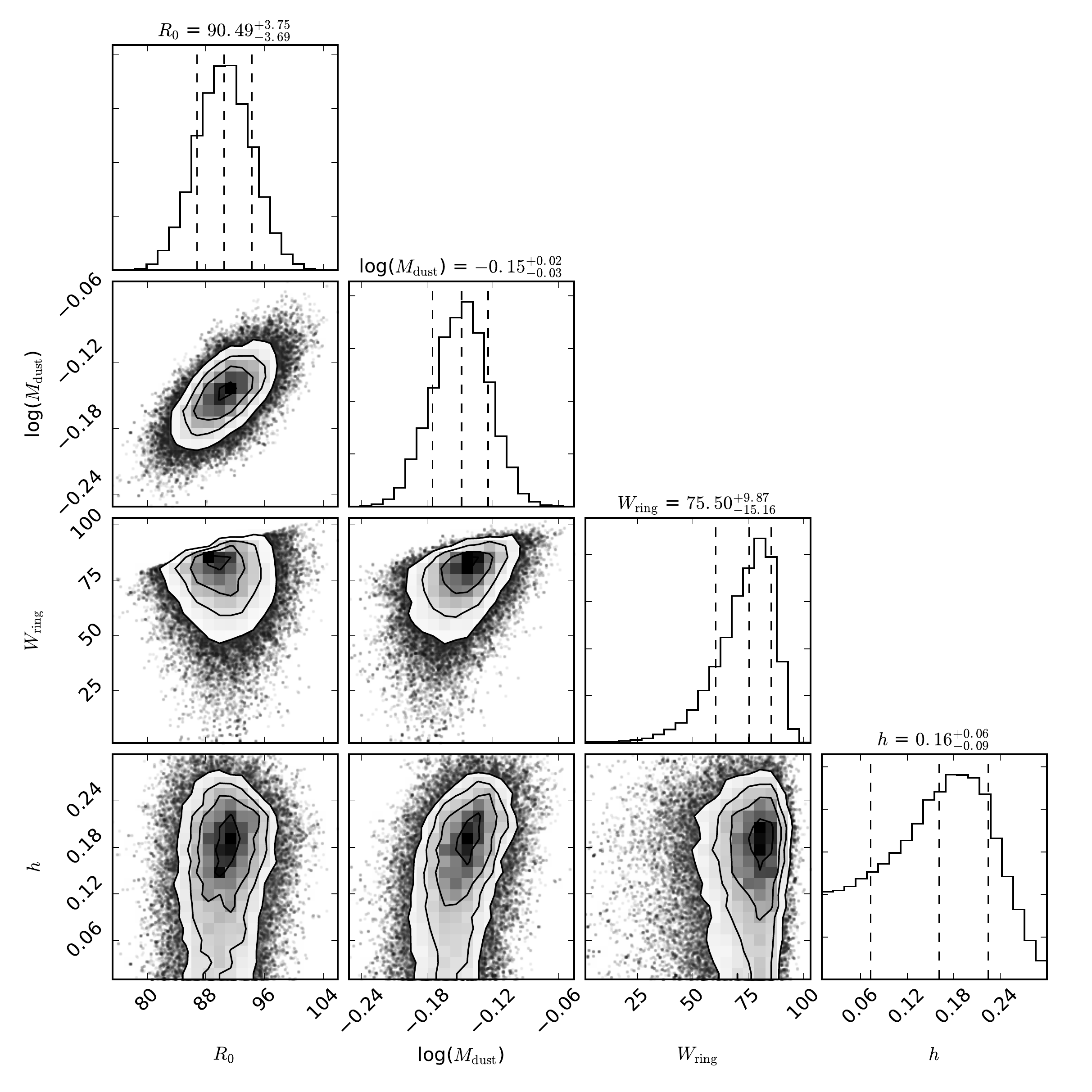}
   \caption{\label{figcornercont} Posterior distribution of the dust continuum model for $R_0$, $M_{\rm dust}$, $W_{\rm ring}$, $h$ (see Sec.~\ref{dustfit}). The marginalised distributions are presented in the diagonal. The vertical dashed lines represent the 16$^{\rm th}$, 50$^{\rm th}$ and 84$^{\rm th}$ percentiles.}
\end{figure*}

We now compare the ALMA visibilities ($V_{\rm obs}$) for the dust continuum to an axisymmetric dust model. The disc model is parametrized as a ring of radius $R_0$ with a Gaussian radial profile of full width at half maximum (FWHM) $W_{\rm ring}$. We also assume a Gaussian
vertical profile of aspect ratio $h=H/R$ ($H$ being the scaleheight) such that the dust density distribution is

\begin{equation}\label{dustmodel}
 \rho_d(R,Z)=\rho_0 \exp\left(-\frac{(R-R_0)^2}{2\sigma_d^2}\right) \exp\left(-\frac{Z^2}{2H^2} \right),
\end{equation}

\noindent where $\rho_0$ is the density at $R_0$ in the midplane and $\sigma_d=W_{\rm ring}/(2 \sqrt{2 \log 2})$.
We then use the line radiative transfer code RADMC-3D \citep{2012ascl.soft02015D} to compute images for a given dust model \citep[as in][]{2016MNRAS.460.2933M} and GALARIO \citep{2018MNRAS.tmp..402T} to convert them into model visibilities ($V_{\rm mod}$) that can be compared to the data in an MCMC fashion.
For the dust optical properties, we assume an astrosilicate composition with a density of 2.7 g$\,$cm$^{-3}$ \citep{2003ApJ...598.1017D} and use a mass-weighted mean opacity of $\kappa=2.5$ cm$^2$g$^{-1}$ at 610 $\mu$m that is computed using the Mie theory code from \citet{1983asls.book.....B}. 
The size distribution is assumed to be between a minimum size of 2.5 $\mu$m \citep[corresponding to the blow out size derived from][]{1979Icar...40....1B} and a maximum size of 1 cm (larger grains do not participate to flux observed in band 8) with a power 
law index of -3.5, which is similar to what is predicted analytically for debris discs \citep{1969JGR....74.2531D} or from numerical simulations \citep[e.g.,][]{2006A&A...455..509K,2007A&A...472..169T,2013A&A...558A.121K}.

\begin{figure}
   \centering
   \includegraphics[width=8cm]{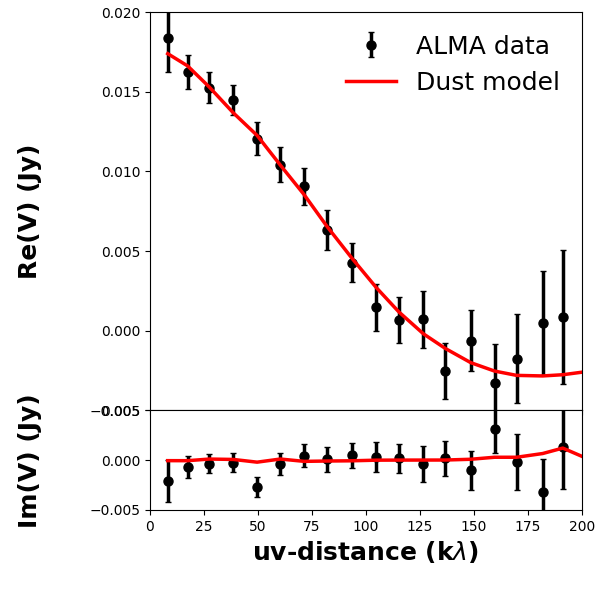}
   \caption{\label{figuv} Real part (top) and Imaginary part (bottom) of the continuum data visibilities (black dots with errors bars) along with the best-fit model (overplotted red line) described in Sec.~\ref{dustfit} and whose parameters are given in Table~\ref{tab1}. The uv-distances are given in units
of observing wavelength and visibilities in Jansky.}
\end{figure}

\begin{table}
  \centering
  \caption{Table describing the best-fit parameters of the dust modelling using an MCMC method (see Sec.~\ref{dustfit}). We list the median $\pm$ uncertainties, which are based on the 16$^{\rm th}$ and 84$^{\rm th}$ percentiles of the marginalised distributions.}

  \label{tab1}
  \begin{tabular}{|l|c|}
   \toprule
    Parameters & Best-fit values \\ 
		 
    \midrule
    \midrule
    $R_0$ (au) & $90.5^{+3.8}_{-3.7}$ \\
    $M_{\rm dust}$ (M$_\oplus$) & $0.71^{+0.03}_{-0.05}$  \\
    $W_{\rm ring}$ (au) & $75.5^{+9.9}_{-15.2}$ \\
    $h$ & $0.16^{+0.06}_{-0.09}$ \\
    $i$ (deg) & $79.4^{+6.8}_{-5.2}$ \\
    PA (deg) & $59^{+2}_{-1.9}$ \\
    RA offset ('') & $-0.09^{+0.02}_{-0.02}$ \\
    Dec offset ('') & $-0.07^{+0.02}_{-0.02}$ \\

   \bottomrule
  \end{tabular}
\end{table}

Therefore, the free parameters that are left, which are fitted in our model, are: $R_0, W_{\rm ring}, h, {\rm inclination}\,(i)$, PA and $M_{\rm dust}$, where the latter is the total dust mass up to 1 cm bodies. We also allow an offset in RA (offset x) and Dec (offset y) to account for astrometric uncertainties in the ALMA data.
We use a Bayesian MCMC approach to constrain the 8 free parameters of the model. We sample the parameter space by using the emcee module \citep[see][for the details of the method]{2010CAMCS...5...65G,2013PASP..125..306F}. We assume that the priors are uniform and the posterior distribution is then given by the
product between the prior distribution function and the likelihood function, which is assumed to be $\propto \exp(-\chi^2/2)$, with $\chi^2= \sum (V_{\rm obs}-V_{mod})^2/\sigma_V^2$ ($\sigma_V$ being the variance of the data). We ran the MCMC with 100 walkers and for 1200 steps after the burn-in period.

The posterior distributions obtained are shown in Fig.~\ref{figcornercont} for the four most important parameters and we summarise all the best-fit parameters in Table~\ref{tab1}. The inclination ($\sim$ 79 deg) and PA ($\sim$ 59 deg) found by our MCMC calculations are consistent with other studies \citep{2015ApJ...815L..14H,2015ApJ...814...42M}.
The offsets in x and y are also consistent with the ALMA astrometric uncertainties ($\sim$0.1'', see ALMA technical handbook\footnote{\url{https://almascience.eso.org/documents-and-tools/cycle6/alma-technical-handbook}}).

The disc radius $R_0$ is constrained to be around 90 au. The width (FWHM) of the ring that we derive from our model is $\sim 76$ au (0.52''), which is however smaller than the beam size 
and thus must be understood as having large uncertainties. A strict upper limit of 95 au can, however, be derived at the 99.7\% level. From these results, 
the bulk of the parent belt would be between $\sim$50 and 140 au.
\citet{2015ApJ...802..138H} observed HD 131835 with T-ReCS in the mid-IR and show that the best fit is for a continuous disc extending from $\sim$40 au to 350 au plus two rings at $\sim$124 and $\sim$260 au (after correcting for the new GAIA distance). The two rings may be made of very small micron-sized grains
(see discussion) and are thus not expected to be seen in the ALMA image. The continuous disc observed with T-ReCS extends farther away than ALMA is sensitive to, because mid-IR observations are sensitive to smaller grains that have very eccentric orbits (owing to radiation pressure) and create an extended halo (whilst ALMA is not).
\citet{2017A&A...601A...7F} observed the disc with SPHERE and detected the presence of 3 concentric rings at $\sim 116, 78,$ and $46$ au. They also published an ALMA band 6 continuum image that looks similar to Fig.~\ref{figcont} with a lower resolution and also find that the ALMA emission is more compact
than at smaller wavelengths.

The dust mass derived from our model is $\sim$0.7 M$_\oplus$. This is consistent with the fit by \citet{2015ApJ...814...42M} who found $0.65\pm0.25$ M$_\oplus$ (after correcting for the new GAIA distance).
These estimates depend on the assumptions made for the composition and size distribution of the bodies whereas the structure does not.
The aspect ratio estimation $h$ is limited because of the relatively low resolution of the observation but we find a best-fit of 0.16 and that it is smaller than 0.29 at 99.7\% confidence level. It, however, may be much smaller than 0.16 as 
observed with SPHERE in the near-IR \citep{2017A&A...601A...7F}.

Finally, we show the best-fit model along with the data in the visibility plane (azimuthally averaged assuming $i=79.4$ degrees and PA = 59 degrees) in Fig.~\ref{figuv} \citep[in a similar way as in][]{2011ApJ...740...38H}. We also subtracted our best fit model from the data in the image plane and checked that the remaining residuals were all within the expected noise level.

\section{ALMA [CI] observation of HD 131835}\label{gas}
\subsection{C$^0$ gas detection}

\begin{figure*}
   \centering
   \includegraphics[width=12cm]{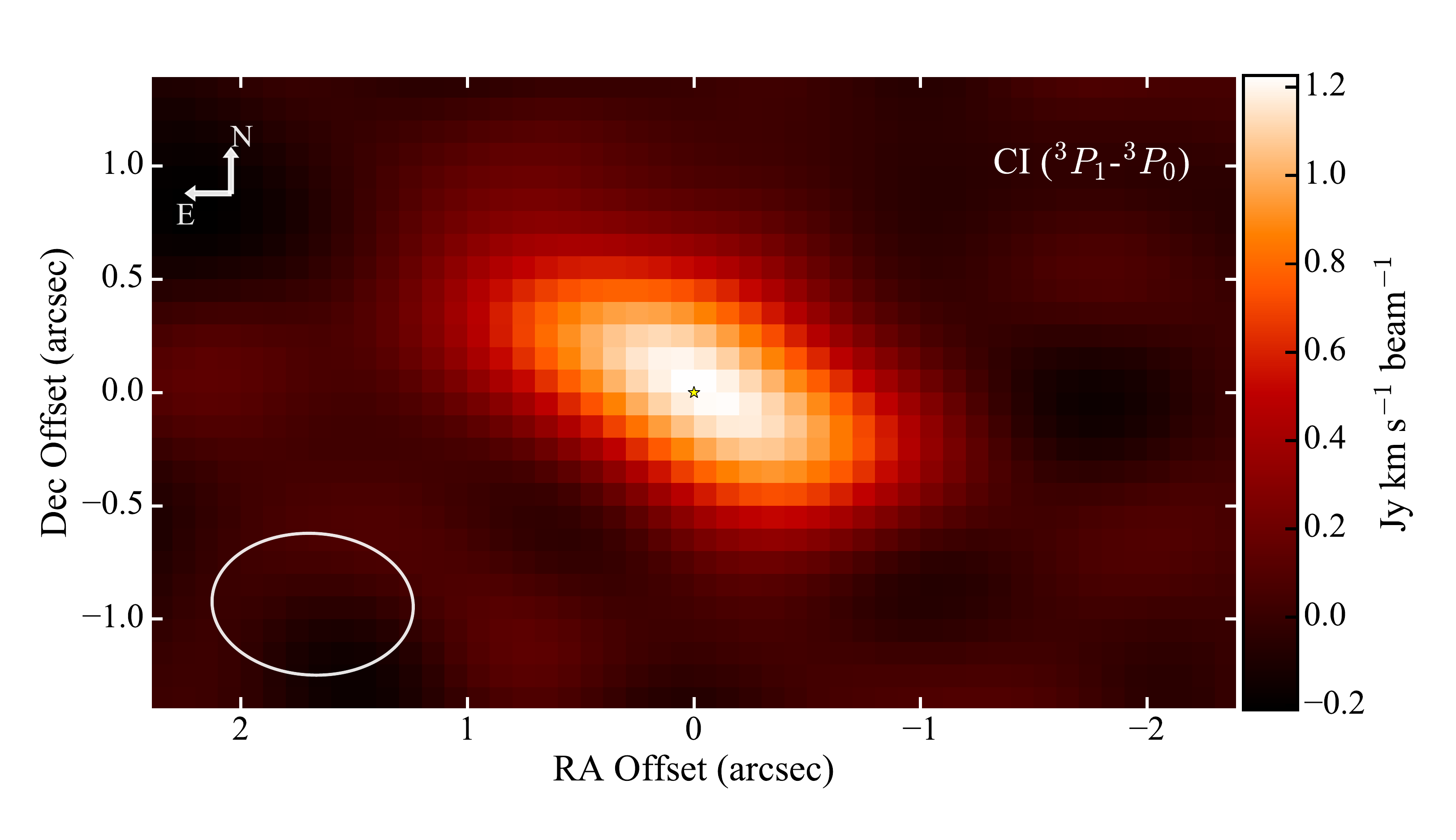}
   \caption{\label{figc1} CI $^{3}$P$_0$-$^{3}$P$_1$ moment-0 (i.e. spectrally integrated) Briggs-weighted (robust=0.5) CLEAN image of HD 131835 (in Jy km/s/beam).
The x-axis shows the RA offset (in arcsec) from the star (shown by a yellow star symbol) and the y-axis the Dec offset (in arcsec). The ellipse shows the beam of 0.89 $\times$ 0.63 arcsec and the North and East directions are shown by the two arrows.
The pixel size is 0.1 arcsec. We integrated the emission between -4 and 10km/s to obtain the moment-0 image.}
\end{figure*}

To obtain the CI $^{3}$P$_0$-$^{3}$P$_1$ (at a rest wavelength of 609.135 $\mu$m) emission map, we first subtract the continuum emission directly from the visibilities (with the task uvcontsub in CASA) using only channels where no [CI] emission is expected.
Fig.~\ref{figc1} shows the moment-0 (i.e. spectrally integrated) [CI] detection around HD 131835. Once again, we show the image to familiarise the reader with the qualitative features of the observation but we only use the observed visibilities in the rest of the paper when it comes to fitting models (see Sec.~\ref{gasfit}). We used the CLEAN algorithm to deconvolve the image using Briggs (with robust=0.5). The synthesized beam (shown as an ellipse in Fig.~\ref{figc1}) has a size of 0.89 $\times$ 0.63 arcsec.
We find that the 1$\sigma$ noise level near the disc in the moment-0 image is 66 mJy$\,$km/s/beam so that the peak S/N is 18. The total integrated flux is 2.8$\pm$0.4 Jy$\,$km/s, which was measured by integrating over a region large enough that it contains all disc emission but small enough that it is not affected by noise contamination (an ellipse of size 3 $\times$ 2.5 arcsec). The error takes account of the noise in the image and flux calibration uncertainties that were added in quadrature.

\begin{figure}
   \centering
   \includegraphics[width=8cm]{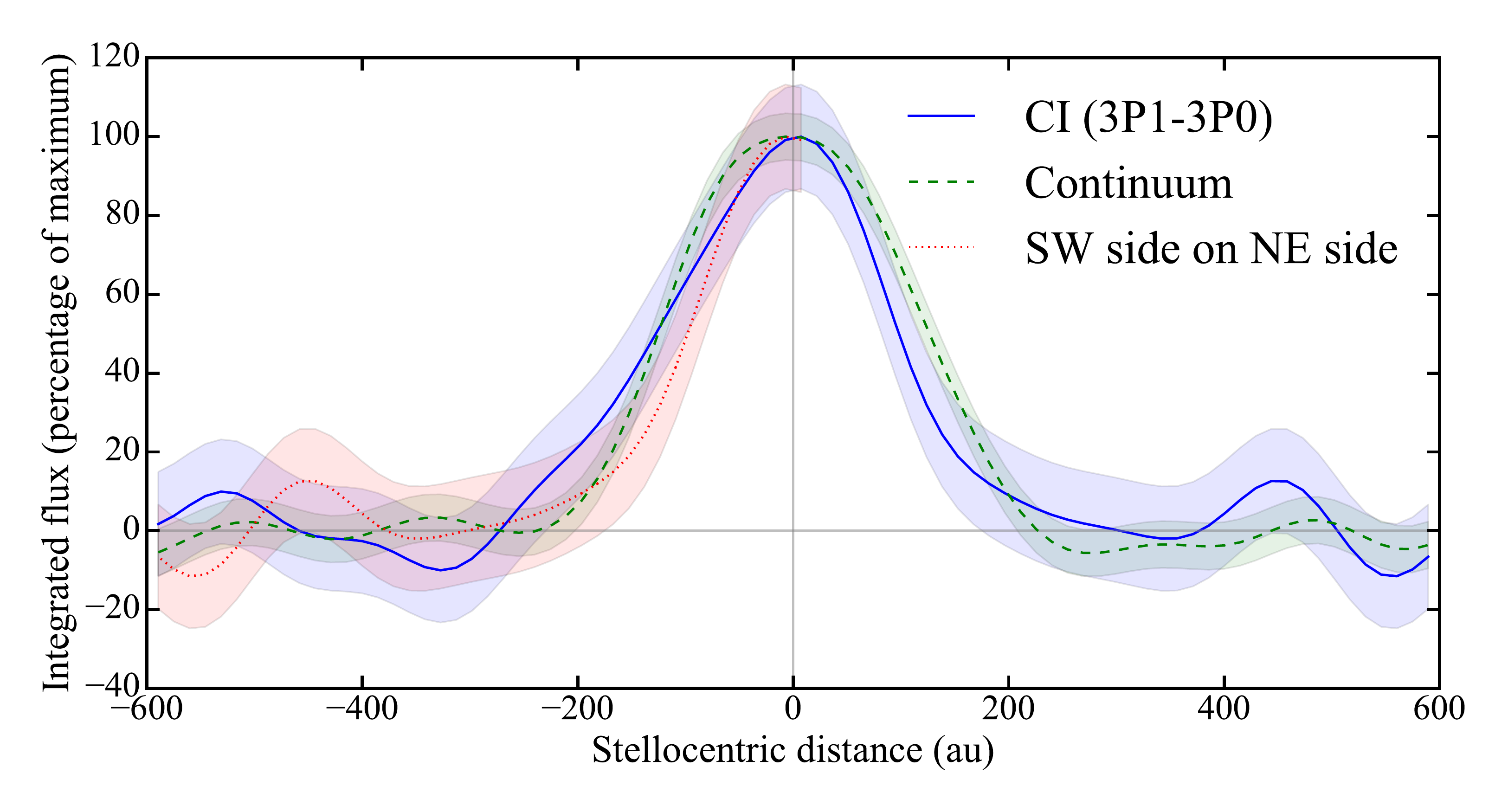}
   \caption{\label{figc1radpro} Radial profile of [CI] (blue solid line) and continuum (green dashed line) obtained by vertically averaging the moment-0 image shown in Fig.~\ref{figc1} and the continuum image from Fig.~\ref{figcont}, respectively. 
We also look for asymmetries in the gas profile by overplotting the SW side of the gas disc on top of the NE side (red dotted line). The errors shown as transparent shaded area are 2$\sigma$. The profiles are normalised by their respective maximum.}
\end{figure}

In Fig.~\ref{figc1radpro}, we show the radial profile of the [CI] emission along the midplane. The flux is integrated vertically within $\pm$100 au from the rotated moment-0 image.
The 2$\sigma$ integrated noise is shown as a blue shaded area around the emission profile. To look for any significant asymmetries, we overplotted the SW side of the disc on its NE side (red dotted line). 
This shows that there are no obvious signs of asymmetry within the allowed error bars. This plot also confirms that most of the emission is within 200 au. We also compare the gas disc emission to the continuum emission (green dashed line) and find
that they are similar. 

\begin{figure}
   \centering
   \includegraphics[width=8cm]{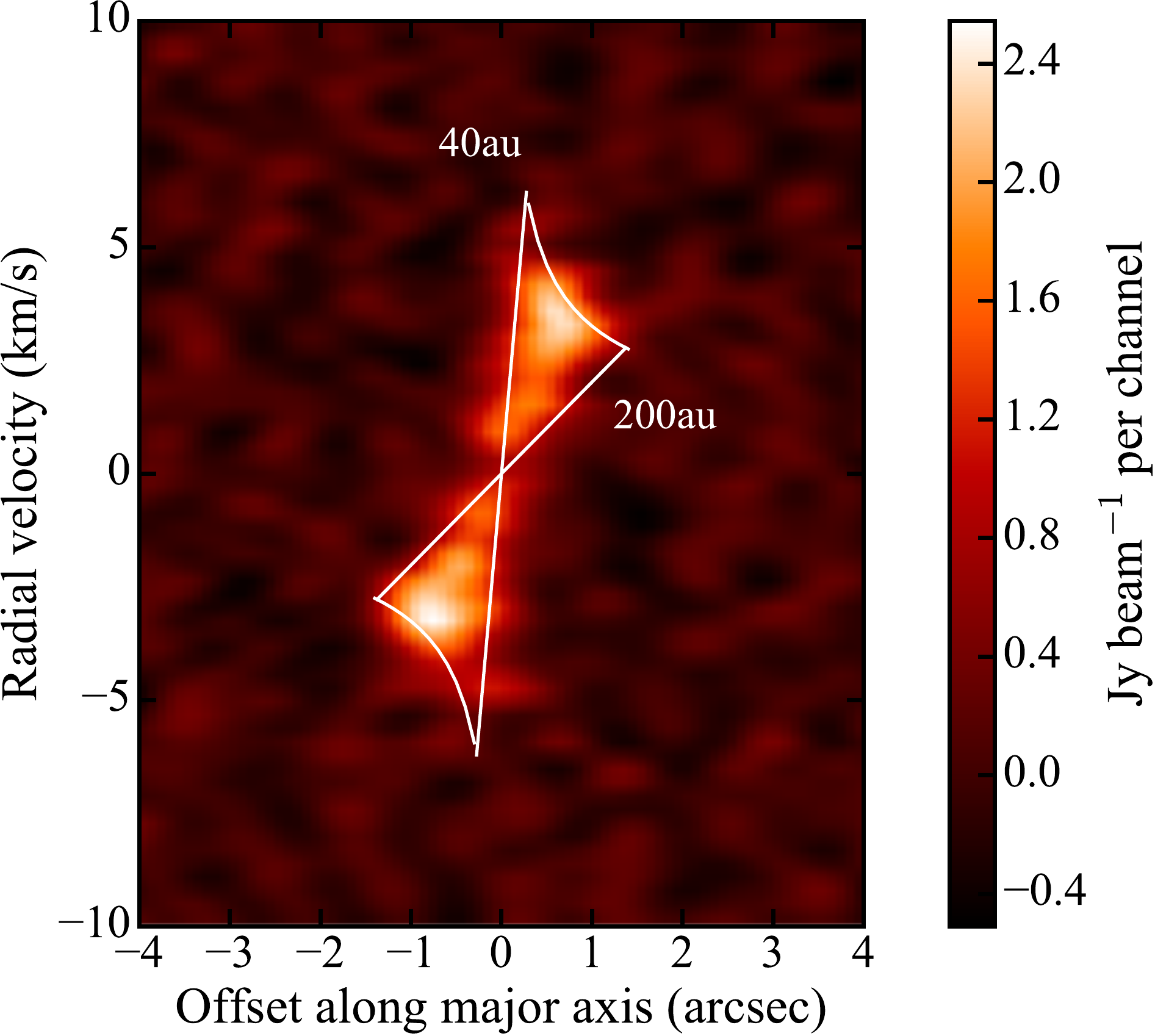}
   \caption{\label{figpv} PV diagram of the [CI] gas disc around HD 131835. The white lines correspond to the locations and radial velocities at which gas is expected to emit if on a circular orbit at 40 and 200 au, respectively.}
\end{figure}

Finally, we have access to the gas kinematics by looking at the emission in the different channels that show the emission at a given radial velocity ($\pm \Delta v$, the channel width). In order to exploit the extra information in velocity it is 
convenient to use a position-velocity (PV) diagram that is shown in Fig.~\ref{figpv}. This is produced by rotating each slice of the [CI] data cube by the PA of the disc and averaging vertically over $\pm$100 au. Each channel then corresponds to a 
horizontal line on the plot. The white lines shown in Fig.~\ref{figpv} correspond to the locations and radial velocities at which gas is expected to emit if 
on a circular orbit at 40 and 200 au \citep[assuming a 1.77 M$_\odot$ star,][]{2015ApJ...814...42M}. The PV-diagram thus tells us that the bulk of the gas disc is between 40 and 200 au and a cavity or a depleted zone may be present interior to 40 au, which we will
test further by comparing to models in the next subsection~\ref{gasfit}. 

\begin{figure}
   \centering
   \includegraphics[width=8cm]{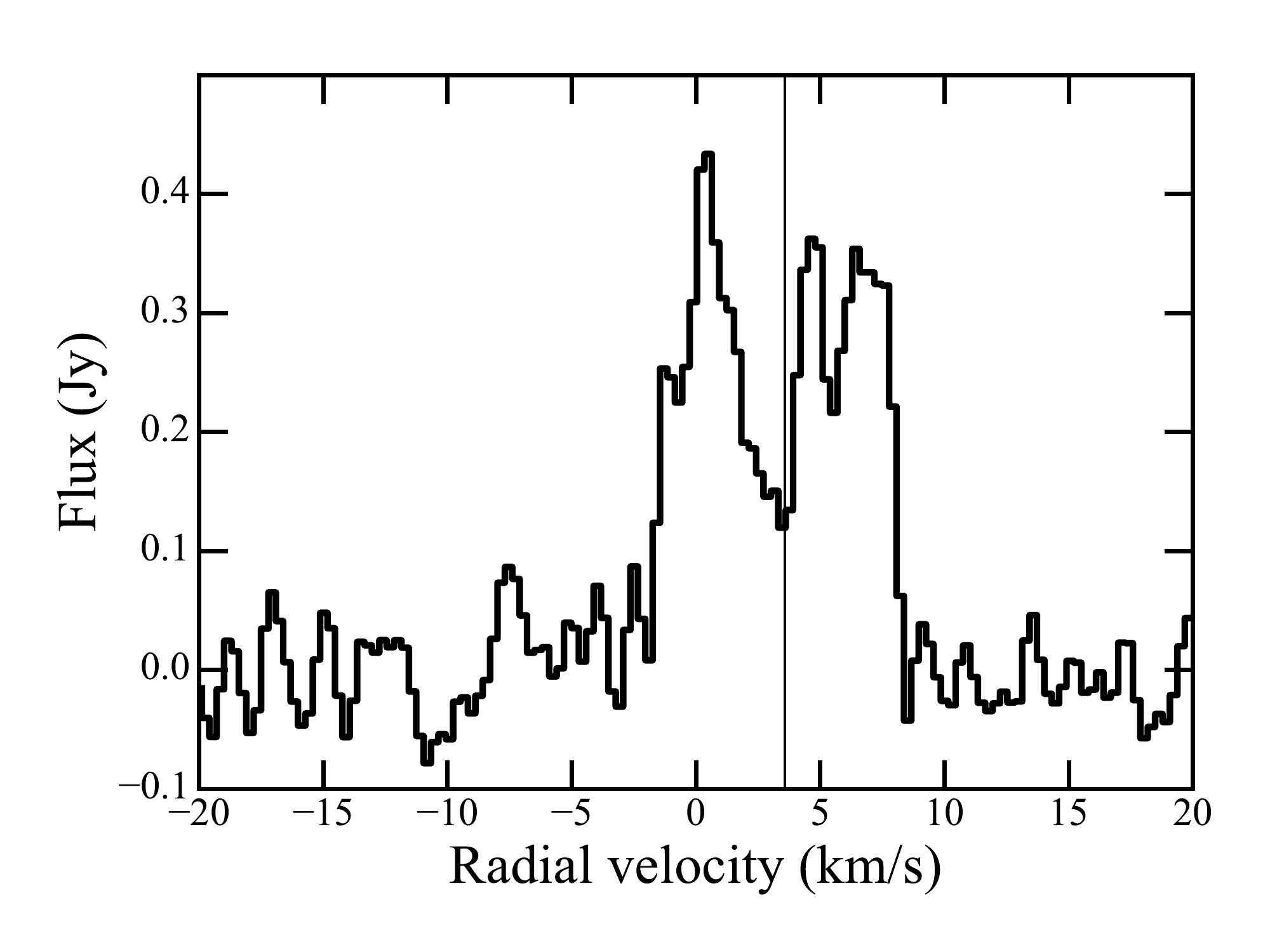}
   \caption{\label{figspec} Spectrum of the [CI] line detected around HD 131835 at a resolution of 0.297 km/s. The thin vertical line corresponds to a velocity of 3.57 km/s.}
\end{figure}

Integrating the PV diagram along its x-axis, we obtain the emission spectrum shown in Fig.~\ref{figspec} at the channel width resolution ($\Delta v=0.297$ km/s). The thin vertical line corresponds to a velocity of 3.57 km/s (which is the best-fit value found from fitting the CI cube, see Sec.~\ref{gasfit}) and is close to the centre of the line.

\subsection{Fit of the [CI] data}\label{gasfit}
To model the gas data, we use the same approach as described in the previous Sec.~\ref{dustfit} for the dust continuum. The only difference being that now we have to produce a cube for the image for the different frequency channels (tracing different radial velocities of the gas).
We also use RADMC-3D to produce the [CI] gas image and convert to visibilities using GALARIO \citep{2018MNRAS.tmp..402T} that can then be compared to the ALMA visibility data cube in an MCMC fashion.

We note that the [CI] emission predicted could be in the non-local thermal equilibrium (non-LTE) regime and thus we would have to add the collider density as an additional free parameter and take account of the radiation from the dust impinging on the gas. We first used the radiative transfer code
LIME \citep{2010A&A...523A..25B} that can handle non-LTE situations to check whether LTE is a good assumption for our case. We find that a good fit of the [CI] line only happens for C$^0$ masses that are high enough that LTE is a good assumption. Indeed, taking the worst case scenario 
which is that electrons are only formed through carbon photoionisation and not from other metals, with a low ionisation fraction, down to $10^{-2}$, and are the only colliders, we find that LTE is still a good approximation. In Sec.~\ref{dali}, we give estimates of the ionisation fraction in HD 131835 from a photodissociation 
region code and show that the ionisation fraction depends on the exact content in carbon and it can vary from $10^{-5}$ to $10^{-1}$.  
However, we find that including hydrogen collisions coming from H$_2$O photodissociation (if water is also released together with CO) would mean that LTE is reached even in the absence of electrons (as then hydrogen atoms can play the role of colliders). 
Indeed, the collision rates for hydrogen and C$^0$ are very similar to those of electrons with C$^0$. For a temperature
between 10 and 200K, the collision rates between H and C$^0$ are $\sim 1.7 \times 10^{-10}$ cm$^3$/s, which translates to a critical H density of $\sim 300$ cm$^{-3}$ to be in LTE. 
This hydrogen density is small compared to what is expected in shielded discs, even for the case where only a small fraction of hydrogen is released together with CO (see Sec.~\ref{comcomp}). We thus assume that LTE is a good approximation even for very low carbon ionisation fractions. 

\begin{figure}
   \centering
   \includegraphics[width=8cm]{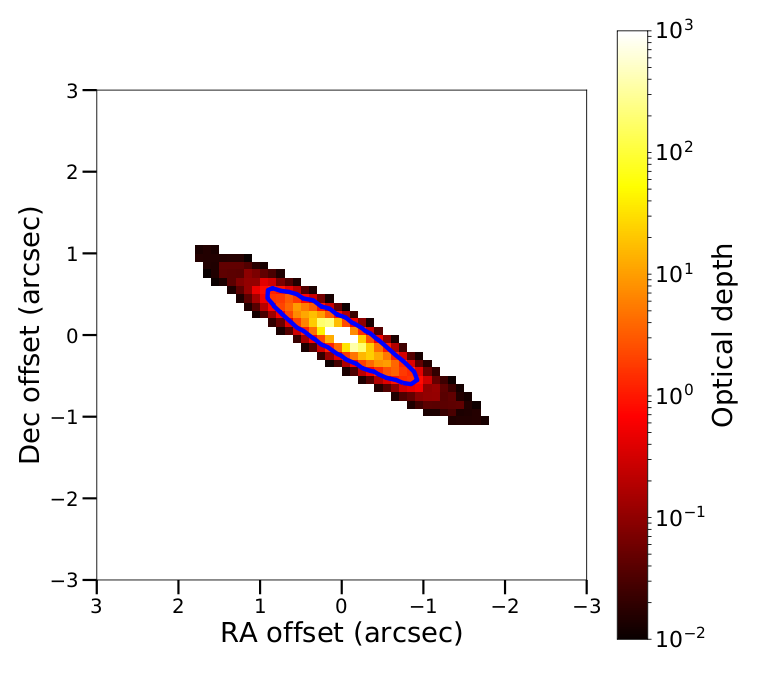}
   \caption{\label{figtau} [CI] optical depth of the best-fit model. The blue contour shows the $\tau=1$ level.}
\end{figure}

After investigation with LIME (and RADMC-3D), we also find that the [CI] line is optically thick for all the cases that match the observed data with an optical thickness $>1$ within 1'', where most of the emission is observed (see Fig.~\ref{figtau}). This means that we will be able to find a good constraint on the gas temperature from the line fitting.
As one LIME model takes about 10 mins to compute, it would anyway be impossible to study a large parameter space in non-LTE and we instead use RADMC-3D in LTE that can compute image cubes in a few seconds and is thus suitable for an MCMC exploration. 

We start by fitting for the inclination $i$, PA and offsets to check that it is similar to the dust disc parameters (see Fig.~\ref{figcorngasi}). We also fit for the star radial velocity, which has been shown to be between 2-4 km/s (heliocentric frame) by previous studies \citep{2006ApJ...644..525M,2015ApJ...814...42M}.
By comparing the posterior distributions of the gas and dust disc, we find that the inclination and PA of both discs are very similar. The astrometric offsets are also consistent between the two data sets. The best fit is for a stellar radial velocity around 3.57$\pm$0.1 km/s (heliocentric frame), which is consistent
with what has been found by previous studies \citep{2006ApJ...644..525M,2015ApJ...814...42M} and shows that the gas is comoving with the star, as expected.

\subsubsection{Gaussian fit}\label{gaussfit}

\begin{figure*}
   \centering
   \includegraphics[width=14cm]{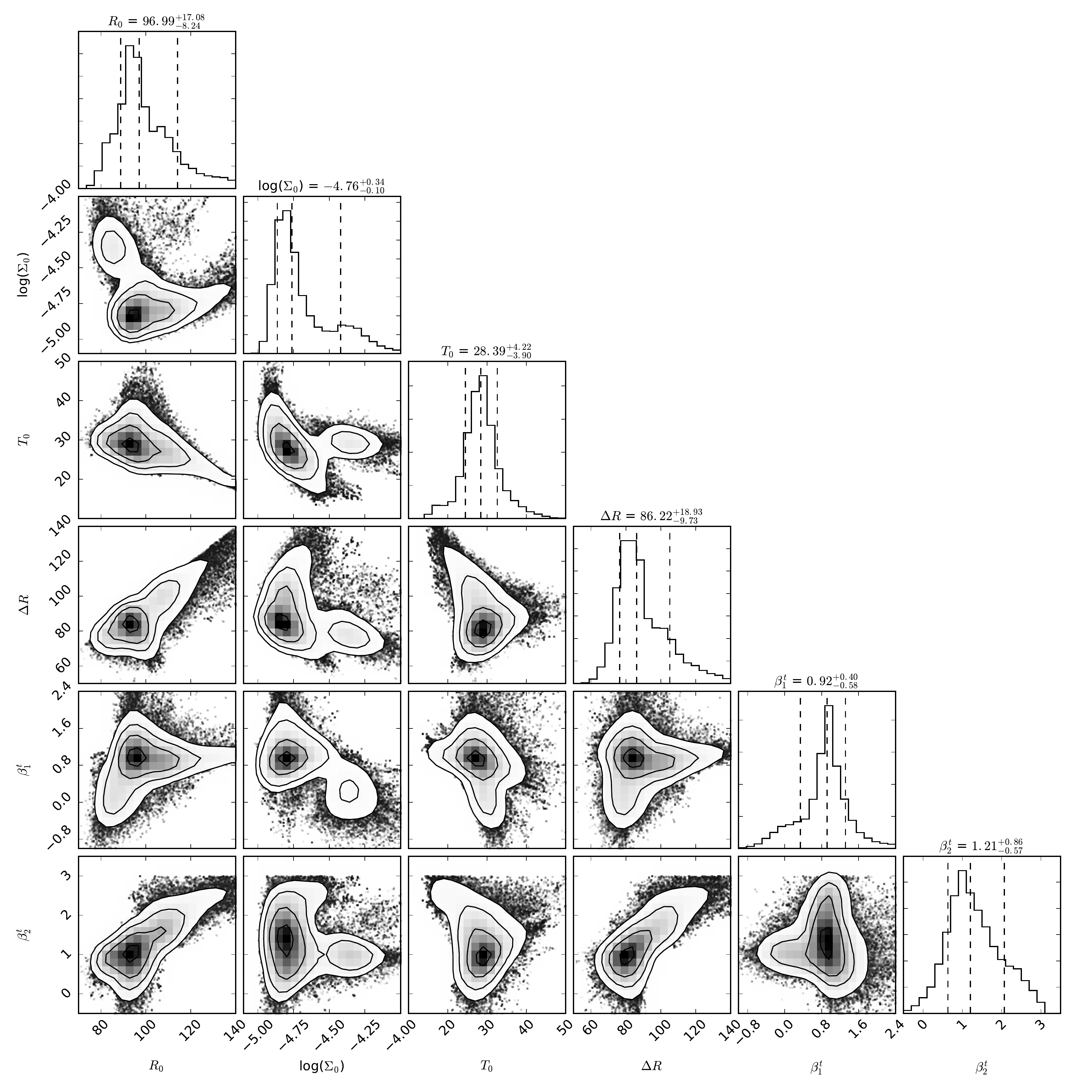}
   \caption{\label{figcorngasg} Posterior distribution of the Gaussian gas model for $R_0$, $\Sigma_0$, $T_0$, $\Delta R$, $\beta_1^t$, $\beta_2^t$ (see Sec.~\ref{gaussfit}). The marginalised distributions are presented in the diagonal. The vertical dashed lines represent the 16$^{\rm th}$, 50$^{\rm th}$ and 84$^{\rm th}$ percentiles.}
\end{figure*}

As the data does not show any hint of asymmetry, we assume an axisymmetric gas model. We first fit the data with a ring having a Gaussian radial profile as for the dust. Therefore, the surface density profile of the gas follows

\begin{equation}
   \Sigma(R) = \Sigma_0  \exp\left(-\frac{(R-R_0)^2}{2\sigma_g^2}\right) ,     
\end{equation}

\noindent where $\Sigma_0$ is the surface density at $R_0$ and $\sigma_g=\Delta R/(2 \sqrt{2 \log 2})$, $\Delta R$ being the width (FWHM) of the gas disc. 
The scaleheight is fixed by the temperature such that $H=c_s/\Omega$, with the sound speed $c_s=\sqrt{kT/(\mu m_H)}$ and $\Omega$ the orbital frequency.
The mean molecular mass $\mu$ could be much higher than in protoplanetary discs as the gas mass
would be dominated by carbon and oxygen in H$_2$-depleted secondary gas discs. We have thus chosen $\mu=14$ as in \citet{2017MNRAS.469..521K}.
The temperature profile is assumed to be a double power law profile \citep[motivated from][]{2016MNRAS.461..845K} defined by

\begin{equation}\label{Teq}
   T(R) =
    \begin{cases}
      T_0 \left(\frac{R}{R_0}\right)^{-\beta_1^t} & \text{for $R<R_0$}\\
      T_0 \left(\frac{R}{R_0}\right)^{-\beta_2^t} & \text{for $R>R_0$}\\
    \end{cases}       
\end{equation}

\noindent The free parameters of this gas model are thus $\Sigma_0$, $\Delta R$, $T_0$, $\beta^t_1$, $\beta^t_2$. The results are shown in Fig.~\ref{figcorngasg} and the best-fit parameters are listed in Table~\ref{tab2}. 

For this Gaussian radial profile, we find that the location of the gas disc $R_0 \sim 97^{+17}_{-8}$ au is consistent with that of the dust disc ($\sim 90$ au). The FWHM of the gas disc is best-fit by $86^{+18}_{-10}$ au, which is slightly larger than that derived for the dust disc ($\sim$ $75^{+10}_{-15}$ au) but is still consistent with being of the same radial extension within error bars. 

The temperature at $R_0$ is best fit by $T_0 \sim 28$ K and a -1 radial power law. 
The best-fit surface density of C$^0$ at $R_0$ is $1.7 \times 10^{-5}$ kg/m$^2$, which corresponds to a column density of $\sim 10^{17}$ cm$^{-2}$, which is enough to start shielding CO (as seen in Sec.~\ref{shield}). This can also be translated into
a number density $\Sigma_0/(2 H \mu m_H) \sim 10^{3}$ cm$^{-3}$, assuming the best fit $T_0$ of $\sim 28$ K to derive $H$. 

\subsubsection{Double power law fit}\label{lawfit}

In the traditional unshielded secondary gas scenario picture, CO is expected to be colocated with the dust, because that CO is released from planetesimals that are traced by the dust seen in the ALMA continuum observations and photodissociates quickly, but this is not necessarily the case for shielded discs (as then CO can viscously spread, see Sec.~\ref{refined}). This is also not necessarily the case for carbon (neutral or ionised) that is not destroyed by photo-chemical processes and can evolve for longer, and may even spread viscously towards the central star
as suggested by \citet{2013ApJ...762..114X,2016MNRAS.461..845K}. To investigate the plausible larger extent of C$^0$, we now assume a double power law radial profile for the gas density splitting at $R_0$. This model is motivated by the expected viscous evolution of C$^0$ (see Fig.~\ref{figevol}). To be consistent, we also use the double power law temperature profile splitting at $R_0$ described by Eq.~\ref{Teq}.
We define the surface density at $R_0$ to be $\Sigma_0$.
The gas surface density is thus given by

\begin{equation}
   \Sigma(R) =
    \begin{cases}
      \Sigma_0 \left(\frac{R}{R_0}\right)^{-\beta_1^s} & \text{for $R<R_0$}\\
      \Sigma_0 \left(\frac{R}{R_0}\right)^{-\beta_2^s} & \text{for $R>R_0$}\\
    \end{cases}       
\end{equation}

\noindent and the temperature profile follows Eq.~\ref{Teq}. 

\begin{figure*}
   \centering
   \includegraphics[width=18cm]{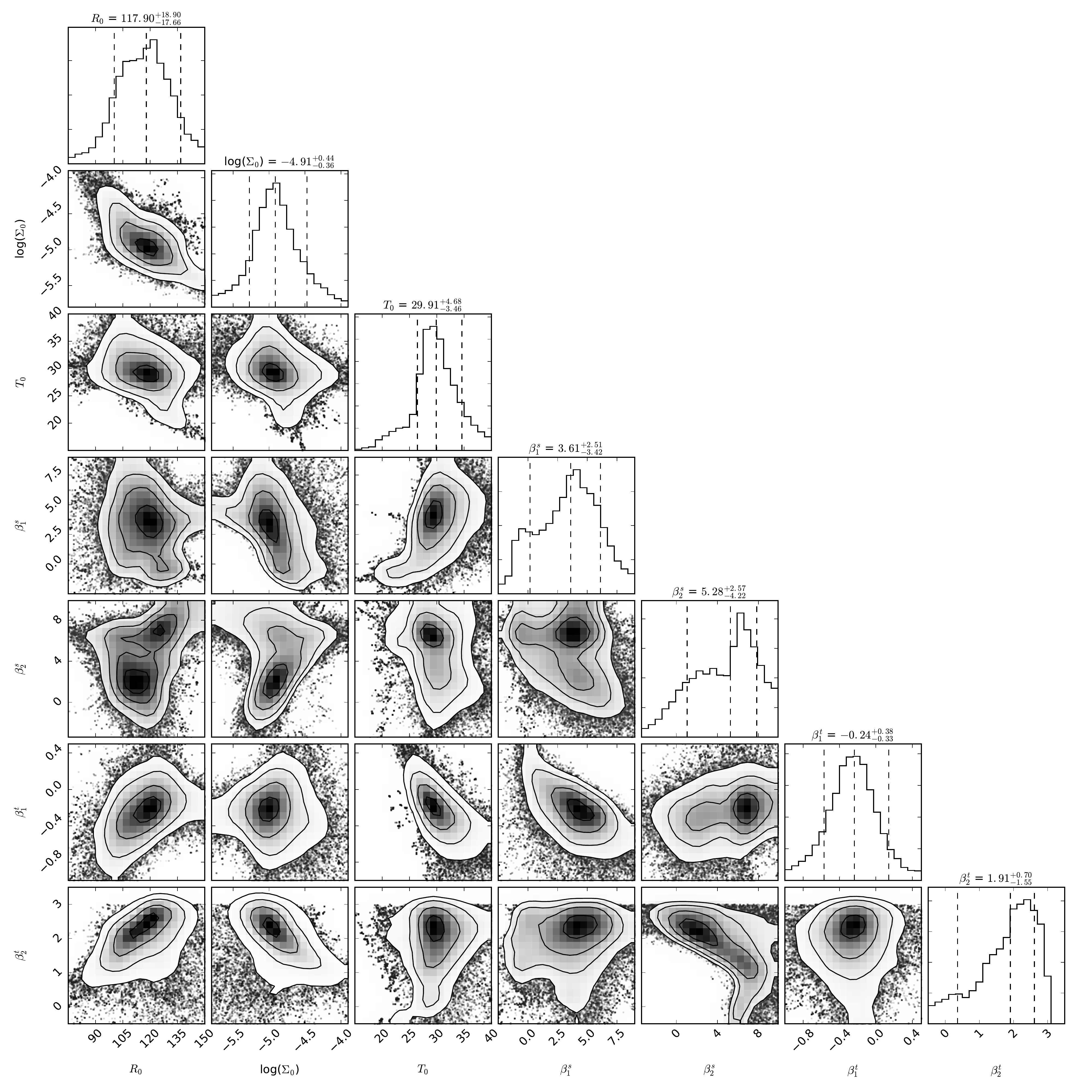}
   \caption{\label{figcorngas} Posterior distribution of the double power law gas model for $R_0$, $\Sigma_0$, $T_0$, $\beta_1^s$, $\beta_2^s$, $\beta_1^t$, $\beta_2^t$ (see Sec.~\ref{lawfit}). The marginalised distributions are presented in the diagonal. The vertical dashed lines represent the 16$^{\rm th}$, 50$^{\rm th}$ and 84$^{\rm th}$ percentiles.}
\end{figure*}

This leaves us with 10 free parameters that are $R_0$, $\Sigma_0$, $T_0$, $\beta_1^s$ $\beta_2^s$, $\beta_1^t$, $\beta_2^t$, $i$, PA and $v_\star$. We explore the parameter space with the same MCMC procedure as presented for the continuum and previous 
gas Gaussian profile and the results are shown in Fig.~\ref{figcorngas}. The best-fit parameters are also summarised in Table~\ref{tab2}.

\begin{table}
  \centering
  \caption{Table describing the best-fit parameters of the Gaussian and double power law [CI] gas modelling using an MCMC method (see Sec.~\ref{gasfit}). We list the median $\pm$ uncertainties, which are based on the 16$^{\rm th}$ and 84$^{\rm th}$ percentiles of the marginalised distributions.}

  \label{tab2}
  \begin{tabular}{|l|c|c|}
   \toprule
    Parameters & \multicolumn{2}{c}{Best-fit values} \\ 
	 & Gaussian & Double power law \\ 

    \midrule
    \midrule
    $R_0$ (au) & $97^{+17}_{-8}$ & $118^{+19}_{-18}$ \\
    $\Sigma_0/10^{-5}$ (kg/m$^2$) & $1.7^{+2.1}_{-0.4}$ & $1.2^{+2.2}_{-0.7}$  \\
    $T_0$ (K) & $28^{+4}_{-4}$ & $30^{+5}_{-4}$ \\
    $\Delta R$ (au) & $86^{+18}_{-10}$ & -  \\
    $\beta_1^s$ & - & $3.6^{+3}_{-3}$ \\
    $\beta_2^s$ & - & $5.3^{+3}_{-4}$ \\
    $\beta_1^t$ & $0.9^{+0.4}_{-0.6}$ & $-0.2^{+0.4}_{-0.3}$  \\
    $\beta_2^t$ & $1.2^{+0.9}_{-0.6}$ & $1.9^{+0.7}_{-1.6}$  \\
    $i$ (deg) & - & $77^{+3.1}_{-2.4}$ \\
    PA (deg) & - & $59^{+1}_{-1}$ \\
    $v_\star$ (km/s) & - & $3.57^{+0.1}_{-0.1}$ \\

   \bottomrule
  \end{tabular}
\end{table}

We find that the gas disc is likely to be peaking in surface density at $R_0 \sim 118$ au but the error bars are large ($\pm 19$ au) because of the low spatial resolution and due to a slight degeneracy with $\Sigma_0$, the surface density at $R_0$. Indeed, if an optically thick gas disc is moved at a greater
distance from the star but with a smaller surface density, the integrated flux over the beam can still be the same as the original disc. 
The best fit surface density at $R_0$ is $1.2^{+2.2}_{-0.7} \times 10^{-5}$ kg/m$^2$, which corresponds to a column density of $\sim 6^{+11}_{-3} \times 10^{16}$ cm$^{-2}$, which is enough to start shielding CO (as seen in Sec.~\ref{shield}). This can also be translated into
a number density $\Sigma_0/(2 H \mu m_H) \sim 0.6^{+1.1}_{-0.3} \times 10^{3}$ cm$^{-3}$, assuming the best fit $T_0$ of $\sim 30$ K and $R_0\sim 110$ au to derive $H$. 

The slope of the surface density
for $R>R_0$, i.e. $\beta_2^s$, is not well constrained but is best fit by a steep value close to 5. This means that there may not be much mass hiding beyond $R_0$. Despite the uncertainties, the probable steepness of the slope is however helpful to constrain the models
as it suggests that the gas disc has not reached steady state. Indeed, the outer surface density slope might be too steep compared to a steady state decretion profile. At steady state, we expect (see Eq.~\ref{Sig}) the accretion profile ($R<R_0$) surface density to scale as $R^{-n}$
and the decretion profile ($R>R_0$) as $R^{-n-1/2}$, where $n$ is the radial scaling in viscosity, i.e., $\nu \propto R^n$ \citep{2012MNRAS.423..505M}. For a temperature profile in $R^\beta$, as $n=1.5-\beta$, it means that for $\beta=1/2$, $\Sigma \propto R^{-1}$ for $R<R_0$
and $\Sigma \propto R^{-3/2}$ for $R>R_0$. If we assume that $\beta_2^t=3/2$, the steady state surface density decretion profile should scale as $R^{-1/2}$, therefore if a profile as steep as $\Sigma(R>R_0) \propto R^{-5}$ is confirmed, this would imply that steady state is not reached, which is most likely the case here (see also section \ref{shield}). But we note that within the error bars, we cannot entirely rule out a steady state profile. Only higher resolution observations would settle this matter.

For $R<R_0$, $\beta_1^s$ is also not well constrained, except to be $>-2.5$ (at the 99.7\% level). This is because the line is becoming optically thick quite rapidly and thus adding more mass in the inner region does not increase the emergent intensity. 
It is therefore hard to tell from the inner accretion profile whether the gas disc is at steady state for which higher spatial resolution observations would again be necessary.

We have also run another MCMC simulation where we added another free parameter $R_{\rm min}$ (on top of the double power law model) below which the surface density becomes zero. We find that an upper limit of $R_{\rm min}=35$ au can be set, meaning that gas is needed to extend at least to 35 au to explain the data. However,
values of $R_{\rm min}<35$ au are not ruled out by the fitting process and it may well be that the gas disc extends further inwards (an observation at higher resolution would also be needed to constrain that better).
The gas disc is thus at a minimum spanning the range 35-100 au, which is slightly more extended (taking into account error bars for the dust model) than the dust disc in the inner regions and could as well be in the outer regions. This favours
that the gas had time to viscously spread since it has been produced. 

The temperature $T$ seems to be well constrained to around 30 K ($\pm4$ K) in the whole disc (at least for $R<R_0$ where $\beta_1^t$ is close to zero). A good constraint arises 
because the line is optically thick (see Fig.~\ref{figtau}) and the observed emission depends only on the temperature and the surface area of emission, which is fixed by the geometry of the disc. In the outer regions, the temperature drops with $\beta_2^t>0$.



\begin{figure}
   \centering
   \includegraphics[width=8cm]{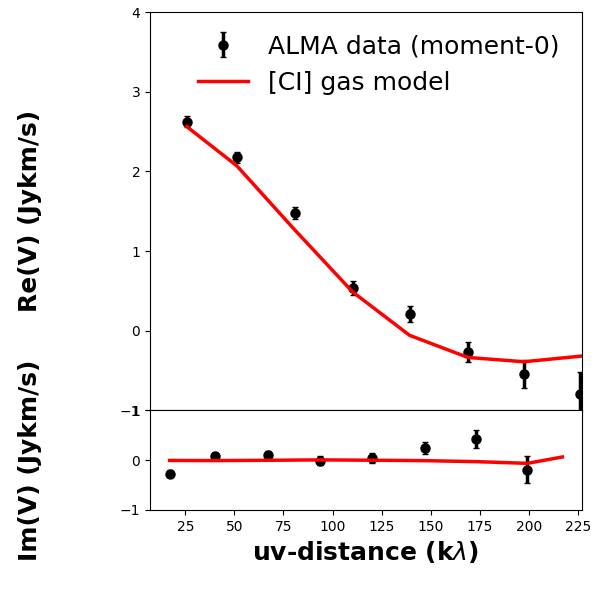}
   \caption{\label{figuvgas} Real part (top) and Imaginary part (bottom) of the [CI] gas moment-0 (integrated across the different channels) visibilities (black dots with errors bars) along with the best-fit model (overplotted red line) described in Sec.~\ref{gasfit} and whose parameters are given in Table~\ref{tab2}. The uv-distances are given in units
of observing wavelength and moment-0 visibilities in Jansky km/s.}
\end{figure}

We compare the best-fit model to the observations in the visibility space by plotting the moment-0 visibilities in Fig.~\ref{figuvgas} (i.e., the visibilities in each channel are summed together), even though the best-fit was found by fitting each of these different channels. The model represents well the overall shape of the visibilities, but leaves a few significant residuals probably linked to the detailed structure of the emission. 

This physical modelling of the [CI] data allows us to place some constraints on the neutral carbon mass around HD 131835. Indeed, we find that the best fit is for a mass $M_{{\rm C}^0}=3.1 \times 10^{-3}$ M$_\oplus$. From the MCMC, we find 
that  $2.7 \times 10^{-3}<M_{{\rm C}^0}<1.2 \times 10^{-2}$ M$_\oplus$ (16$^{\rm th}$ and 84$^{\rm th}$ percentiles), assuming that $\beta_1^s<3$. Higher values of
$\beta_1^s$ would be unphysical because the steady state level for the range of temperature slopes $\beta=[-1,0]$ derived from the MCMC are expected to 
be in $R^{\beta-3/2}$ (see Eq.~\ref{Sig}). 
The C$^0$ mass is not well constrained because the line is optically thick in the whole disc but we will show in Sec.~\ref{app} that in order to match the CO
mass derived from an ALMA optically thin line observation \citep{2017ApJ...849..123M} with this C$^0$ mass imposes a narrower range of C$^0$ masses. We note that the C$^0$ mass derived here is in the right range of values to be able to start shielding CO from photodissociating as can be seen from 
Fig.~\ref{figCIneed} presented in the introduction of shielded discs (see Sec.~\ref{shieldeddd}).

\section{Discussion}\label{discu}

\subsection{Temperature, ionisation fraction, and chemical reactions expected for a shielded disc of secondary origin}\label{dali}

From our modelling of the [CI] line, we find that $2.7 \times 10^{-3}<M_{{\rm C}^0}<1.2 \times 10^{-2}$ M$_\oplus$. The C$^{+}$ upper limit from Herschel (at 30 K, assuming LTE) is $2\times 10^{-3}$ M$_\oplus$ \citep{2015ApJ...814...42M}. This means that the upper limit on the ionisation fraction
is $\sim 0.4$, which is already lower than that found for $\beta$ Pic \citep{2014A&A...563A..66C,Catal}. We now use the DALI 2D code\footnote{The low gas density and very low abundance of hydrogen in the system we study is strictly outside the parameter range for which DALI -- a protoplanetary disc code -- is validated to give correct results. We caution the reader to interpret our results as a guide to future work rather than a final result for the HD 131835 disc.} \citep{2012A&A...541A..91B,2013A&A...559A..46B} to compute the carbon ionisation fraction in the disc taking into account the UV flux from the central star and IRF. 
For the DALI simulations, we start with a cometary abundance of hydrogen, carbon and oxygen (i.e., for a carbon abundance set to 1, then $N_{\rm H}=20$ and $N_{\rm O}=11$). We set a star spectrum similar to HD 131835, i.e. a 15 Myr A-type star of about 10 L$_\odot$.
We impose that the total carbon mass is around $10^{-2}$ M$_\oplus$ and has a constant surface density between 1 and 100 au (with an aspect ratio of 0.1),
which is consistent (within error bars) with values derived from the [CI] observations.
Each simulation takes a day to compute and we clearly cannot iterate to a best-fit model but we use the simulations to have a feel for the typical ionisation fraction and temperature to expect in these shielded discs.

\begin{figure*}
   \centering
   \includegraphics[width=8cm]{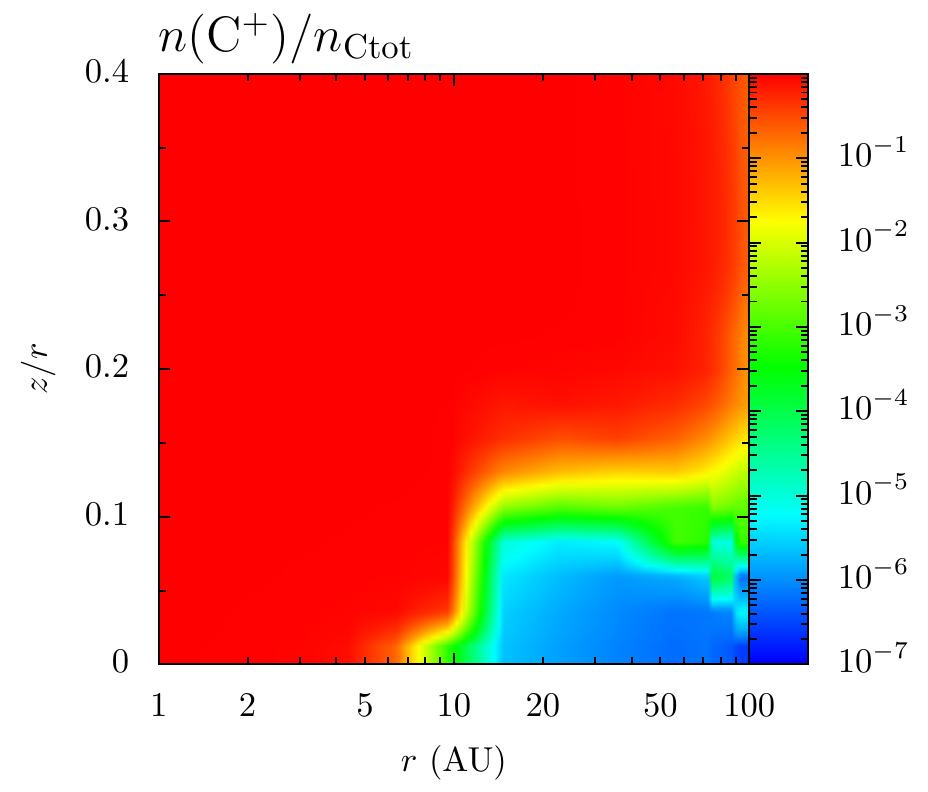}
   \includegraphics[width=8cm]{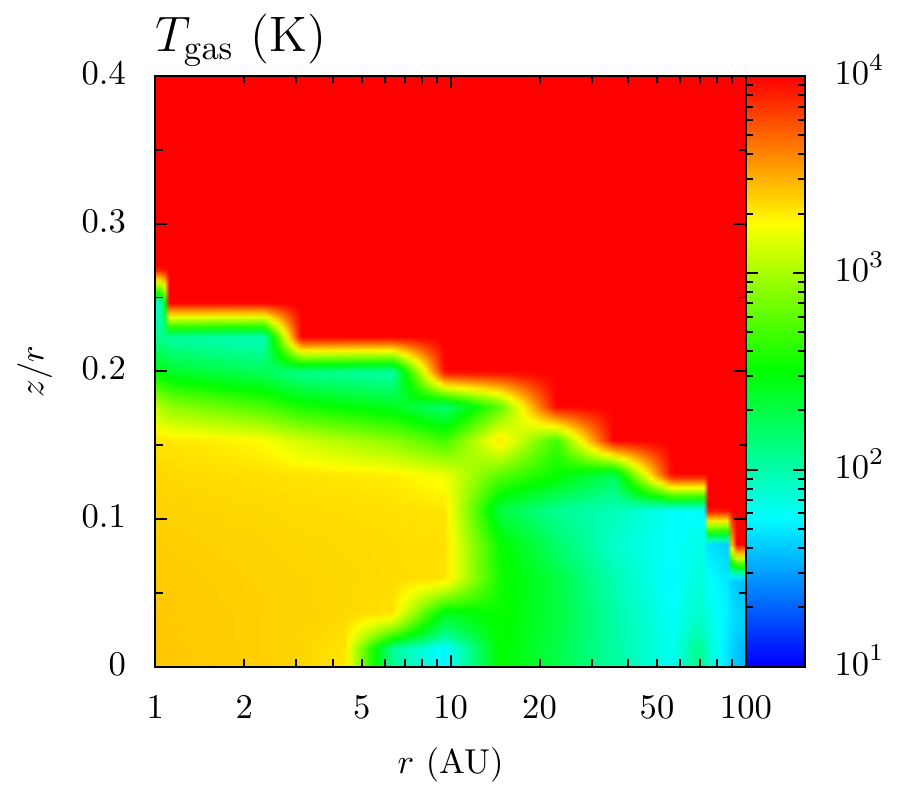}
   \caption{\label{figionftdali} {\it Left:} Ionisation fraction of carbon in a shielded disc simulated with the DALI code. $z$ is the vertical direction in the disc. {\it Right:} Gas temperature in a shielded disc simulated with the DALI code.}
\end{figure*}

Fig.~\ref{figionftdali} (left) shows the result of the simulation for the ionisation fraction. We expect that in the midplane of the disc and beyond 10 au, the ionisation fraction can become very small ($\sim 10^{-5}$) owing to C$^0$ shielding that catches most of the UV photons from the IRF in the upper
layer so that the midplane cannot be ionised. Within 10 au, the ionisation fraction goes back up because of the UV photons coming from the star. We note that our grid started at 1 au and we expect that this is an effect of the grid because the radial optical depth would be much higher for a grid 
starting at the star location and therefore UV photons from the star would be caught at a much smaller radius, thus decreasing the ionisation fraction below 10 au. Our assumption that the ionisation fraction is low in these discs, lower than 0.1, seems to be justified from this test simulation.
We also note that we ran some simulations with 10 times less carbon and the ionisation fraction was becoming higher, closer to 0.1. It is therefore dependent on the exact carbon mass at different radii.
Higher resolution images would enable us to pinpoint the carbon density as a function of radius and compute a better estimate of the ionisation fraction in the disc.


Fig.~\ref{figionftdali} (right) shows the simulation results for the temperature in a typical shielded disc (i.e., with enough carbon to shield CO over long timescales). We find that, indeed, the temperature can reach values of order 30-40K in the midplane close to 100 au. This is in agreement with the temperature obtained from fitting the optically thick [CI] line in Sec.~\ref{gasfit}.
We note that the temperature can indeed be lower than the dust temperature (which would be around 50K at 100 au around a star similar to HD 131835 assuming black body emission). By looking at what is setting the temperature, we find that the dominant heating is from C$^0$ photoionisation and the dominant
cooling is from the OI and CII lines.

CO is not the only molecule that can be shielded owing to C$^0$.
In the DALI simulations, we find that CN, N$_2$ as well as CH$^{+}$ are also shielded by C$^0$ and thus, they can accumulate over time. 
This effect can be quantified using the results from the study by \citet{2012MNRAS.427.2328R}. Similar to the photodissociation timescale of CO
used in Eq.~\ref{tco}, \citet{2012MNRAS.427.2328R} find that C$^0$ shielding increases the photodissociation timescales of CN, N$_2$ and CH$^{+}$
as $1/(S+(1-S)\exp(-\sigma_i N_{{\rm C}^0}))$. Whilst $S=0$ in Eq.~\ref{tco} for CO, it is equal to 0.018, 0, and 0.036 for CN, N$_2$ and CH$^{+}$, respectively. This means that for these 3 molecular species, C$^0$ shielding can be almost as strong as for CO.
We thus predict that the photodissociation timescales of these species, respectively 60, 190 and 100 yr 
\citep[assuming photons coming from the IRF,][]{1988ASSL..146...49V} could be orders of magnitude longer than expected 
in unshielded discs \citep[see][]{2018ApJ...853..147M}, which may lead
to these species being detectable in shielded discs together with C$^0$. CH$^{+}$ would be particurlarly interesting to detect as it would give some constraints
on the amount of hydrogen in the system (as it would mainly come from C+H) and potentially on the amount of water on the planetesimals.

\subsection{New models to refine the total C$^0$ mass and viscosity ($\alpha$) in the gas disc}\label{CIm}

In this section, we first start by presenting a new analytical model of shielded discs that takes into account the shielding of CO by C$^0$ and shows what range
of $\alpha$ or $\dot{M}_{\rm CO}$ values can lead to a shielded configuration (Sec.~\ref{an}). This analytical model can be useful to understand the different scalings but it lacks accuracy as it neglects CO self-shielding. That is why we then develop a more sophisticated semi-analytical box model in Sec.~\ref{refined} that takes into account CO self-shielding and shows the temporal evolution of the build-up of gas at the radius $R_0$ of gas injection. We use this more accurate model in Sec.~\ref{app} to fit the observed CO mass in HD 131835 and find what typical $\alpha$ value is needed to explain the data. In turn, it provides the C$^0$ mass that can accumulate for a given $\alpha$ and CO production rate that we can compare to our new CI observations finding that indeed our model is consistent with both CO and CI data.
We note that this refined model is 0D but could be improved by turning it into a 1D model in the future as it would be useful when higher resolution observations are at hand. For the data we have, such a model would not be useful and is therefore deemed beyond the scope of this paper.

\subsubsection{An analytic model}\label{an}
As previously explained in Sec.~\ref{gasfit}, the C$^0$ mass is not well constrained from the [CI] observation we presented but assuming that the gas produced is secondary, we can improve the constraints on the C$^0$ mass using our knowledge of the much more accurate CO mass (since 
this was measured from an optically thin line) and computing the C$^0$ mass that would be required to explain this amount of CO. With the knowledge of this more accurate (model-based) C$^0$ mass, we can then put some constraints on the viscosity (parametrized with $\alpha$) that would allow such a C$^0$ mass to accumulate over the age of the system.
 
Assuming that CO is produced at a rate $\dot{M}_{\rm CO}$ and that the CO mass required to fit the observations is $M_{\rm CO}$, a photodissociation timescale $t_{\rm CO}=M_{\rm CO}/\dot{M}_{\rm CO}$ is required. Using Eq.~\ref{tco}, this translates to the C$^0$ mass needed to shield CO

 \begin{equation}\label{mc1}
 M_{{\rm C}^0{\rm n}}=\frac{2 \pi R \Delta R \mu_c m_H}{\sigma_i} \ln \left(\frac{M_{\rm CO}}{120{\rm yr} \times \dot{M}_{\rm CO}}\right),
\end{equation}

\noindent where $\dot{M}_{\rm CO}$ should be expressed in units of mass per yr and $\mu_c$ is the mean molecular weight of carbon. The C$^0$ mass is set through the CO input rate such that $\dot{M}_{{\rm C}^0}=12/28 (1-f) \dot{M}_{\rm CO}$, where $f$ is the ionisation fraction. C$^0$ at $R_0$ is assumed to evolve to steady state over
local viscous timescales such that to get the C$^0$ mass needed $M_{{\rm C}^0{\rm n}}$, the local viscous timescale should be $t_{\nu}(R_0)=M_{{\rm C}^0{\rm n}}/\dot{M}_{{\rm C}^0}=R_0^2/(12 \nu(R_0))$, where $\nu=\alpha c_s^2/\Omega$ is the viscosity parametrized by an $\alpha$ parameter. This translates as a condition on $\alpha$
to get the C$^0$ mass needed and therefore the CO mass required

 \begin{equation}\label{alphane}
 \alpha=\frac{\mu \sigma_i \sqrt{G M_\star} \dot{M}_{\rm CO} (1-f)}{ 56 \pi \mu_c \Delta R k T_0 \sqrt{R} \ln \left(\frac{M_{\rm CO}}{120{\rm yr} \times \dot{M}_{\rm CO}}\right)}.
\end{equation}

\begin{figure}
   \centering
   \includegraphics[width=8cm]{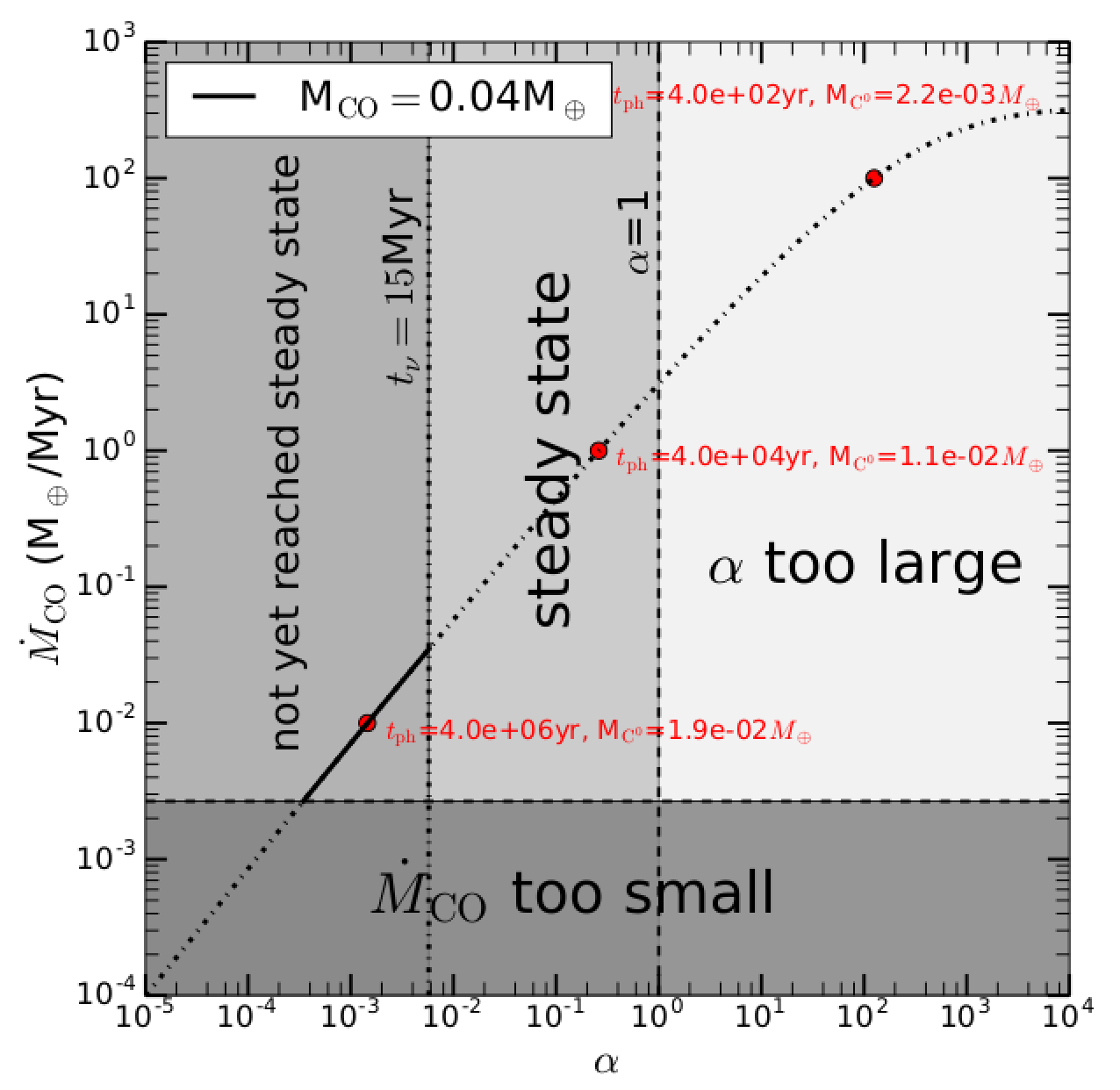}
   \caption{\label{figalpha} The dash-dotted black line shows the $\alpha$ value needed to produce enough C$^0$ that can shield CO and produce a total CO mass of 0.04 M$_\oplus$ as a function of the production rate of CO in the belt $\dot{M}_{\rm CO}$. The two vertical lines
show $\alpha=1$ (dashed), and $\alpha=6\times10^{-3}$ (dotted) corresponding to a viscous timescale of 15 Myr (i.e., the age of HD 131835). The horizontal dashed line shows the minimum value of $\dot{M}_{\rm CO}$ to have been able to produce 0.04 M$_\oplus$ of CO within 15 Myr.
The 3 red points show the corresponding photodissociation timescales and C$^0$ masses for three different $\alpha$ values corresponding to $\dot{M}_{\rm CO}=[10^{-2},1,10^{2}]$M$_\oplus$/Myr, respectively. The solid section of the dash-dotted line shows the $\alpha$ range that can explain the CO and C$^0$ observations within the frame of our analytical model described in Sec.~\ref{an}.}
\end{figure}

Fig.~\ref{figalpha} shows the value of $\alpha$ that is required to produce a C$^0$ mass that can produce the observed 0.04 M$_\oplus$ of CO for different values of $\dot{M}_{\rm CO}$ for a disc located at $R_0=90$ au with, again, a width $\Delta R=70$ au. 
To the right of the black dash dotted line, the CO mass produced would be below 0.04 M$_\oplus$ (because $\alpha$ is so high that there is insufficient C$^0$ to shield CO); similarly more CO would be expected to the left of the line. There are 4 different regimes shown on the plot with different grey shaded areas.
At the extreme right of the plot, $\alpha>1$ and is therefore too large to be physical. In the middle, this is the regime where the gas disc has had time to reach steady state. At the extreme left of the plot, $\alpha$ is small and the gas disc has not yet had time to reach steady state.
The viscous timescale must be $>15$ Myr (age of the system) to not be at steady state, which corresponds to $\alpha<6\times10^{-3}$. 
The horizontal dashed line shows the minimum value of $\dot{M}_{\rm CO}$ to have been able to produce a total CO mass of 0.04 M$_\oplus$ over 15 Myr\footnote{We note that there is also a maximum production rate if we assume
that the whole belt mass, which cannot be more massive than the solid mass available in protoplanetary discs (i.e., $\sim 0.1 \times M_\odot/100 \sim 300$ M$_\oplus$ assuming a gas-to-dust ratio of 100), is released in CO over 15 Myr, which would give $\sim$ 20 M$_\oplus$/Myr.}. 
We also showed in Sec.~\ref{gasfit} that the gas disc is most likely not at steady state as the outer surface density slope may be too steep and thus the disc
could have spread only for a maximum of 15 Myr (likely much less as secondary production of gas presumably starts when the protoplanetary disc is dispersed, i.e. after a few Myr). 

To respect all the constraints so far, we are thus left with the 
part of parameter space shown by the solid line, i.e., $2.5 \times 10^{-4}<\alpha<6 \times 10^{-3}$ and $2 \times 10^{-3}<\dot{M}_{\rm CO}<4 \times 10^{-2}$ M$_\oplus$/Myr. For instance, we calculate the corresponding CO
photodissociation timescale and C$^0$ mass for $\dot{M}_{\rm CO}=10^{-2}$ M$_\oplus$/Myr (the leftmost red point) that are 4 Myr and $1.9\times10^{-2}$ M$_\oplus$, respectively (which is for $\alpha=1.5\times10^{-3}$). 
The C$^0$ mass obtained is thus compatible with the C$^0$ mass derived from observations in Sec.~\ref{gasfit}. We note that in the process of converting $\dot{M}_{\rm CO}$ to a C$^0$ input rate (to then work out the $\alpha$ value needed to explain the observed CO mass), we assumed an ionisation fraction $f$ of 0.1. However,
this does not affect the results because the dependence is in $1-f$ and $f$ is assumed to be $<0.1$ as already discussed in Sec.~\ref{dali}.

The range of $\dot{M}_{\rm CO}$ expected for HD 131835 from a secondary gas production model can be derived from the dust luminosity using Eq.~2 in \citet{2017MNRAS.469..521K} and does not depend on the photodissociation timescale. This calculation finds 
$10^{-2}<\dot{M}_{\rm CO}<0.5$ M$_\oplus$/Myr by assuming a CO-to-dust mass ratio on the comets between 1 and 30\% \citep[i.e., similar to the range observed for Solar System comets,][]{2011ARA&A..49..471M} but we caution that the uncertainties
can be up to a factor 10 on this range \citep{2017MNRAS.469..521K}.
This predicted range and the plausible $\dot{M}_{\rm CO}$  range needed overlap and are thus compatible, which gives confidence that a secondary model can indeed explain the dust, neutral (and ionised) carbon and CO observations all together.

The derived value of $\alpha$ is much smaller than that derived in the $\beta$ Pic gas disc \citep[for which we found $\alpha>0.1$,][]{2016MNRAS.461..845K}. However, this might be expected for the gas disc around HD 131835 owing to the optical thickness to UV photons that can ionise carbon resulting in a much lower ionisation fraction and thus a lower $\alpha$ value if the viscosity is
triggered by the magnetorotational instability \citep[][and references therein]{1998RvMP...70....1B,2016MNRAS.461.1614K}. This low-$\alpha$ value could also come from non-ideal MRI effects as a lower ionisation fraction may result in ambipolar diffusion becoming important, thus slowing down the viscous evolution \citep[see][]{2016MNRAS.461.1614K}.
In any case, this lower value of $\alpha$ is compatible with the wide range of values found in protoplanetary discs \citep[e.g. between $10^{-4}$ and 0.04 as estimated in][]{2017ApJ...837..163R}.

\subsubsection{A more refined semi-analytic model}\label{refined}

\begin{figure*}
   \centering
   \includegraphics[width=17cm]{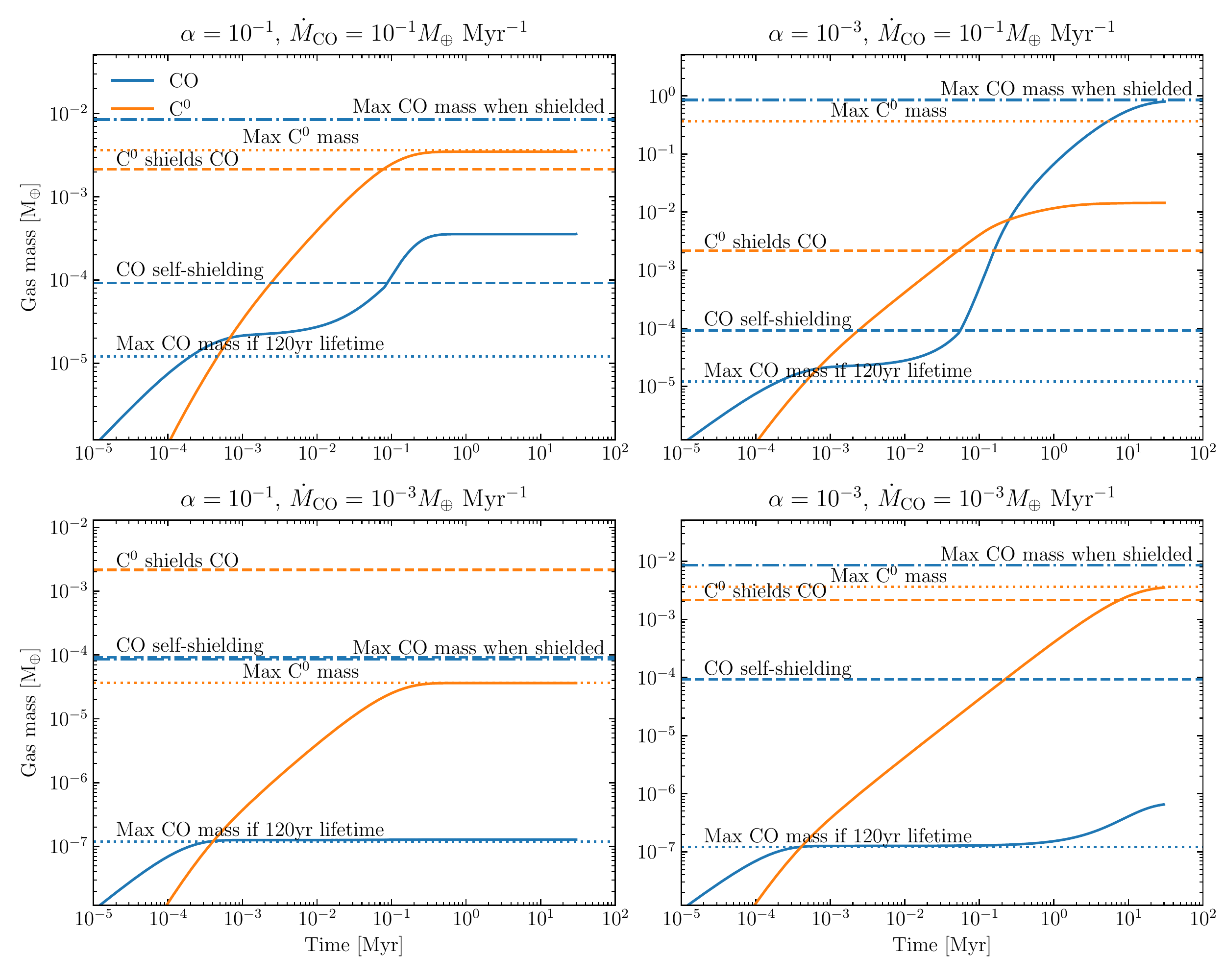}
   \caption{\label{figcoc1evol} CO (blue) and C$^0$ (orange) gas mass evolution as a function of time for different $\dot{M}_{\rm CO}$ ($10^{-3}$, and $10^{-1}$ M$_\oplus$/Myr) and $\alpha$ values ($10^{-3}$, and $10^{-1}$). The dotted horizontal blue line shows the maximum CO mass
that can be reached if the CO lifetime is 120 yr (i.e., no shielding) and the orange one is the maximum C$^0$ mass that can be reached when viscous spreading balances the C$^0$ input rate. The dashed horizontal blue (orange) line shows when CO self-shielding (C$^0$ shielding of CO) becomes important.
The dash-dotted blue line is the maximum CO mass that can be reached when CO is shielded and that the viscous timescale becomes smaller than the photodissociation timescale, i.e. this CO mass is when viscous spreading balances the CO input rate. For these plots, we assumed 
that $R_0=90$ au, $T=30$ K and $\Delta R=70$ au.}
\end{figure*}

The analytic model described in Sec.~\ref{an} is now improved to take into account the temporal evolution of the gas as well as CO self-shielding that can become important for the high CO masses involved. Now, we also include a negative feedback on the total C$^0$ mass, which is that for an increasing C$^0$ mass,
CO has time to viscously spread (because of the longer photodissociation timescale) and thus less C$^0$ is produced from CO at a given radius (which means CO destruction and C$^0$ production cannot be assumed to be constant in time). The previous section should therefore just be used to get an idea of the different scalings but not to compute predictions. 
For the new refined semi-analytic model, we start with no gas at $t=0$ and input a constant CO gas mass $\dot{M}_{\rm CO}$ per unit time. The CO mass $M_{\rm CO}$ is then evolved over time at the radius $R_0$ as

\begin{equation}\label{dMco}
\frac{{\rm d}M_{\rm CO}}{{\rm d}t}=\dot{M}_{\rm CO}^+-\dot{M}_{\rm CO}^-,
\end{equation}

\noindent where $\dot{M}_{\rm CO}^+=\dot{M}_{\rm CO}$ is the production rate of CO and $\dot{M}_{\rm CO}^-=M_{\rm CO}/t_{\rm CO removal}$ is the destruction rate of CO, with 
$t_{\rm CO removal}=(1/(t_{\rm COph})+1/t_\nu)^{-1}$, where $t_{\rm COph}$ includes the photodissociation timescale due to C$^0$ shielding (Eq.~\ref{tco}) and self-shielding \citep[tabulated from][]{2009A&A...503..323V}.

In a similar way, the carbon mass $M_{\rm C}$ follows

\begin{equation}\label{dMc1}
\frac{{\rm d}M_{\rm C}}{{\rm d}t}=\dot{M}_{\rm C}^+-\dot{M}_{\rm C}^-,
\end{equation}

\noindent where $\dot{M}_{\rm C}^+=(12/28) M_{\rm CO}/t_{\rm COph}$ and $\dot{M}_{\rm C}^-=M_{\rm C}/t_\nu$. 

We evolve this set of equations for 30 Myr for different $\alpha$ values ($10^{-1}$ and $10^{-3}$) and different $\dot{M}_{\rm CO}$ ($10^{-1}$ and $10^{-3}$ M$_\oplus$/Myr) and the resulting CO and C$^0$ masses are plotted in Fig.~\ref{figcoc1evol} in blue and orange, respectively. 
Motivated by our previous results, for these plots we assume that $R_0=90$ au, the ionisation fraction is close to 0, the temperature is $\sim 30$ K and the planetesimal belt spreads over $\Delta R=70$ au.
The classical unshielded case, where shielding of CO by C$^0$ is unimportant is shown in the bottom left panel of Fig.~\ref{figcoc1evol}. This happens for high $\alpha$ values or low $\dot{M}_{\rm CO}$. In this case, we see that the steady state CO mass (which 
is $120\,{\rm yr} \times \dot{M}_{\rm CO}=1.2 \times 10^{-7}$ M$_\oplus$ and shown as a blue horizontal dotted line) is reached in less than 500 yr. The C$^0$ mass accumulates until reaching a few viscous timescales (which is $\sim$0.1 Myr for $\alpha=0.1$) 
when the viscous spreading balances the carbon input rate. The final C$^0$ mass reached is $\dot{M}_{{\rm C}^0} t_\nu = 3.6 \times 10^{-5}$ M$_\oplus$ as shown by the horizontal orange dotted line. 

The other case for which $\dot{M}_{\rm CO}=10^{-3}$ M$_\oplus$/Myr (bottom right plot) is very similar
but eventually the C$^0$ mass becomes large enough that C$^0$ starts shielding CO from photodissociating (the C$^0$ shielding mass above which this happens is shown as a horizontal 
orange dashed line, which is taken to be when $\sigma_i N_{{\rm C}^0}=1$ in Eq.~\ref{tco} and is close to $2 \times 10^{-3}$ M$_\oplus$ as shown in Fig.~\ref{figCIneed}). This is why after $\sim 1$ Myr, 
the CO mass goes back up again before plateauing to its new steady-state value. This starts happening because the maximum C$^0$ mass that can be reached is higher than the mass to start shielding CO (i.e., $\dot{M}_{{\rm C}^0} t_\nu>2 \pi R \Delta R \mu_c m_H/\sigma_i$), 
which was not the case previously for a higher $\alpha$. 

\begin{figure}
   \centering
   \includegraphics[width=8cm]{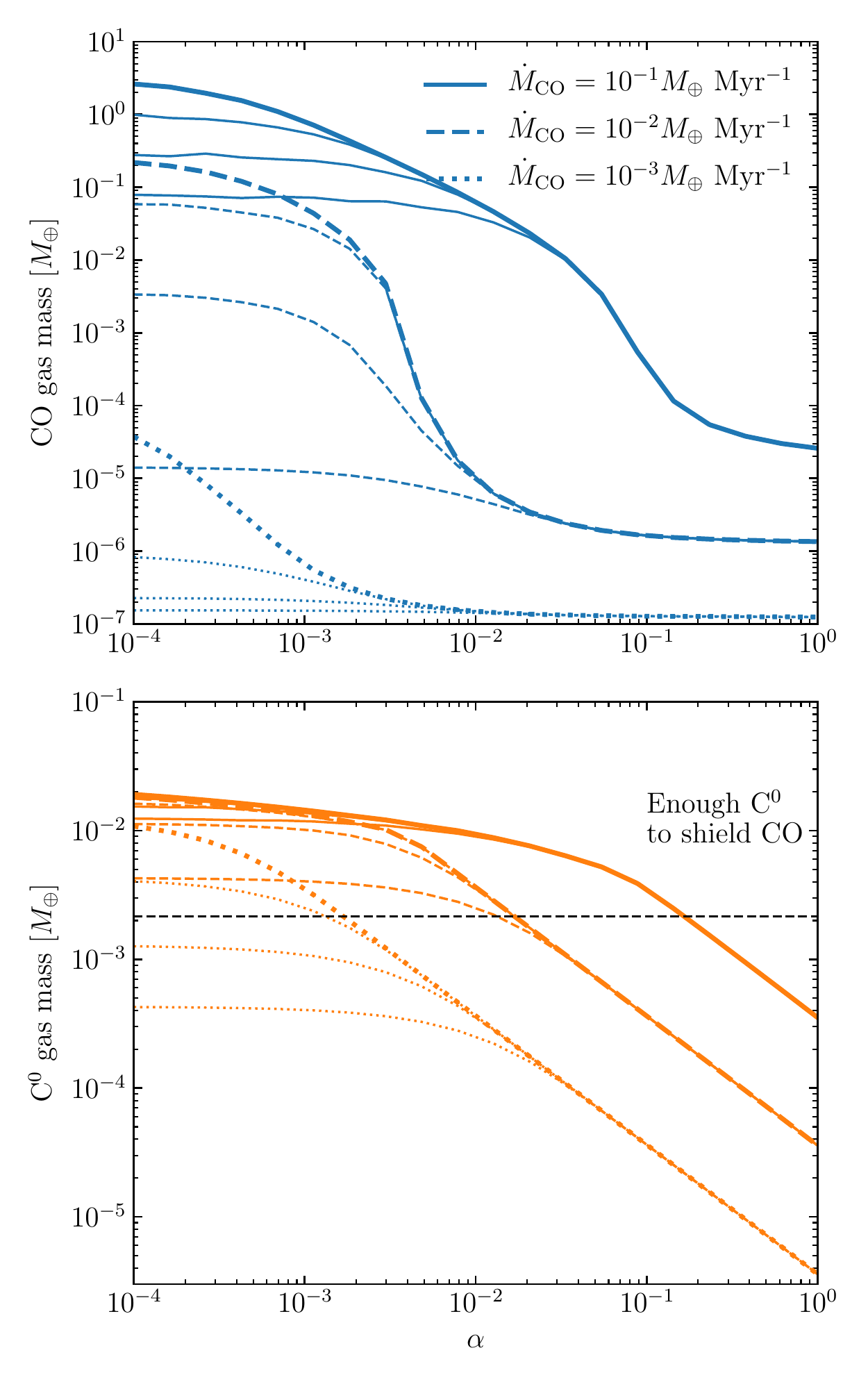}
   \caption{\label{figcoc1alpha} CO (top) and C$^0$ (bottom) gas masses after 1, 3 and 10 Myr (thin lines) and 30 Myr (thick) as a function of $\alpha$ for 3 different $\dot{M}_{\rm CO}$ of $10^{-1}$ (solid), $10^{-2}$ (dashed), and $10^{-3}$ M$_\oplus$/Myr (dotted). This shows the final stages of gas evolution
depicted in Fig.~\ref{figcoc1evol} for a wider variety of $\alpha$ and $\dot{M}_{\rm CO}$. We keep the blue and orange colours for CO and C$^0$, respectively, to match the convention used in Fig.~\ref{figcoc1evol}. 
The horizontal dashed line in the bottom plot shows when the C$^0$ mass becomes high enough to start shielding CO significantly.}
\end{figure}

The top left plot is for a higher CO input rate (and high $\alpha$) and it evolves similarly to the case that has just been presented, except for the intermediate plateau that is now affected by CO self-shielding.
The blue horizontal dashed line at $8 \times 10^{-5}$ M$_\oplus$, is when the CO photodissociation timescale is increased due to self-shielding by $e \sim 2.72$ but self-shielding starts affecting the CO mass before CO reaches that level, and this further increases the CO lifetime.   

The top right subplot in Fig.~\ref{figcoc1evol} is the most characteristic of what could explain shielded discs (note the more extended y-scale for this subplot). In this case, the C$^0$ mass that can be reached is two orders of magnitude above where C$^0$ starts shielding CO, meaning that the exponential term in Eq.~\ref{tco}
can reach very high values and therefore CO can survive several orders of magnitude longer than without shielding. The CO mass that is reached for this case is $\sim 0.8$ M$_\oplus$, i.e. more than 4 orders of magnitude higher than the expected mass of $\sim 10^{-5}$ M$_\oplus$ without shielding.
For this case, shielding from C$^0$ dominates over self-shielding. The shielding timescale becomes so long that
the maximum CO mass that can be reached becomes dominated by viscous spreading, which is why the CO mass plateaus at $\dot{M}_{\rm CO} t_\nu \sim 0.8$ M$_\oplus$ (blue dash-dotted line). The C$^0$ mass does not reach its maximum theoretical value of $\dot{M}_{{\rm C}^0} t_\nu$ (orange dotted line)
because CO spreads more rapidly than it is photodissociated and thus the maximum C$^0$ mass that can be reached is a factor $\sim t_\nu/t_{\rm COph}$ smaller. This also means that as CO spreads viscously, it is not necessarily colocated with the main dust belt (as was originally thought for gas discs of
secondary origin), but it should in any case be colocated with the carbon gas disc (at least in the region where C$^0$ shielding is efficient, i.e., this colocation may not apply at large radii, see Fig.~\ref{figevol}).

Fig.~\ref{figcoc1alpha} shows the CO (top) and C$^0$ (bottom) gas masses after 1, 3, 10 Myr (thin lines) and 30 Myr (thick) as a function of $\alpha$ for 3 different $\dot{M}_{\rm CO}$ of $10^{-1}$ (solid), $10^{-2}$ (dashed), and $10^{-3}$ M$_\oplus$/Myr (dotted). For high $\alpha$ values (or low $\dot{M}_{\rm CO}$), the final CO
mass reached is $120$ yr $\times \, \dot{M}_{\rm CO}$, as expected without shielding and the C$^0$ mass reaches its maximum at a value of 12/28 $\dot{M}_{\rm CO} t_\nu \propto \dot{M}_{\rm CO}/\alpha$, which explains the C$^0$ evolution with $\alpha$ and $\dot{M}_{\rm CO}$.
When $\alpha$ becomes smaller or $\dot{M}_{\rm CO}$ becomes higher, a new regime is attained where the CO mass reaches orders of magnitude higher values which can explain the typical CO masses observed with ALMA for shielded discs. The final CO mass in this regime
plateaus because it is then dominated by the viscous evolution since the photodissociation timescale becomes greater than the viscous timescale. In this shielded regime, the C$^0$ mass plateaus because the CO mass input rate at $R_0$ becomes lower owing to CO spreading and thus
less C$^0$ can be produced per unit of time. 

Therefore, this figure shows that there is clearly two regimes of secondary gas production with a sharp transition between gas discs of low CO and C$^0$ masses (e.g. $\beta$ Pic) and high CO and C$^0$ masses (e.g. HD 131835) depending on $\alpha$ and $\dot{M}_{\rm CO}$. However, we notice
that if the gas production only started recently (i.e., $\sim$1 Myr, lowermost thin lines on Fig.~\ref{figcoc1alpha}) and for low values of $\dot{M}_{\rm CO}$, 
a disc that is on its way to becoming shielded could be mistaken for an unshielded low-mass secondary gas disc. This is because before becoming a high
mass gas disc, there is a build-up process, the timescale of which is fixed by the time to reach a C$^0$ mass high enough to start shielding CO as shown in Fig.~\ref{figcoc1evol}. Also, we note that the gas input rate will decrease over time because the dust mass loss rate will 
become smaller owing to collisions that are depleting the belt mass \citep[e.g.][]{2007ApJ...663..365W,2008ApJ...673.1123L}. Thus, a shielded disc can become unshielded again after a certain period of time.

\subsubsection{Application of the refined semi-analytic model to HD 131835}\label{app}

\begin{figure*}
   \centering
   \includegraphics[width=18cm]{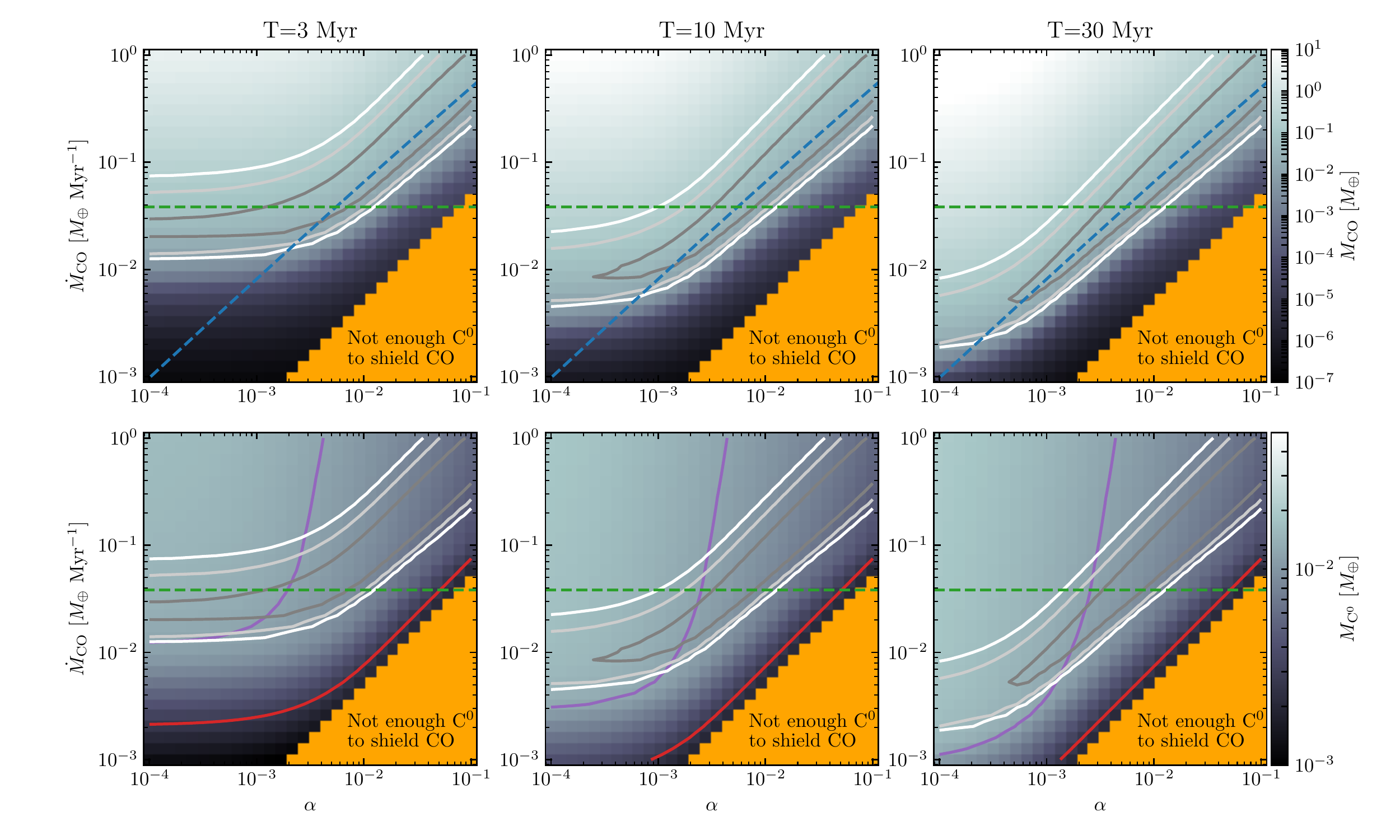}
   \caption{\label{figco13} Best-fit values to get $M_{\rm CO}=0.04$ M$_\oplus$ as observed with ALMA for HD 131835, in the parameter space $\dot{M}_{\rm CO}$ VS $\alpha$. These results are based on the model presented in Sec.~\ref{refined}. The three columns correspond to different evolution time from when the secondary production of gas has started. 
{\it Top:} The colourscale is showing the CO mass $M_{\rm CO}$ and the contours are plotted for 1, 3 and 5$\sigma$ from the observed value. The yellow region is where C$^0$ does not reach densities that are high enough to shield CO. The blue dashed line is the prediction from the analytical model presented in Sec.~\ref{an}. The green dashed line shows the expected CO production rate $\dot{M}_{\rm CO}$ from the stellar and 
disc parameters \citep[see][]{2017ApJ...842....9M}. 
{\it Bottom:} Total C$^0$ mass needed to produce $M_{\rm CO}=0.04$ M$_\oplus$ in different parts of the parameter space. The red and purple lines are for C$^0$ masses of $2.7 \times 10^{-3}$ and $1.2 \times 10^{-2}$ M$_\oplus$ (predicted range from observations), respectively. We overlay the CO contours from the top plots to check where the best-fit models lie in terms of C$^0$ masses.}
\end{figure*}

We now apply the model presented in the previous section \ref{refined} to the specific case of HD 131835 and try to explain the high observed CO mass of 0.04 M$_\oplus$ in this system. We ran a grid of 1000 models for different $\dot{M}_{\rm CO}$ and $\alpha$ values and in Fig.~\ref{figco13} (top), we compute the predicted CO mass shown in colour scale and plot contours that show where the predicted CO mass is within 1, 3, and 5$\sigma_{\rm CO}$ from the mass inferred from the observations, where the uncertainty $\sigma_{\rm CO}$ is assumed to be 0.5 dex from the observed mass \citep{2017ApJ...849..123M}.
As the models are not always at steady state, we ran simulations for different ages. HD 131835 is $\sim$15 Myr old but the secondary production started after the protoplanetary disc dispersed, i.e., likely after a few Myr. We plot the best-fit models for 3, 10 and 30 Myr.
10 Myr is likely to be the most realistic unless the gas production is linked to a recent event (a few Myr old) and we also show 30 Myr when most of the parameter space is then at steady state. The horizontal part of the best-fit region (i.e., where the contours are horizontal for small $\alpha$) is for models that have not reached 
steady state yet. The yellow region shows where the C$^0$ mass produced through the
secondary evolution is not high enough to shield CO from photodissociating (i.e., it is not the shielded disc regime). In the rest of the parameter space, CO is shielded owing to C$^0$ and the contours show where the CO mass is closest to 0.04 M$_\oplus$. The green dashed line shows the CO production rate predicted from the stellar and dust disc parameters \citep[see][noting that uncertainties are up to a factor 10 on that value]{2017ApJ...842....9M}. 
From the 10 Myr plot, we conclude that $10^{-3}<\alpha<10^{-2}$ can explain the observed CO mass, which is similar to what was found with the more simple analytical model presented in Sec.~\ref{an} as can be seen from the blue dashed line (on top plots of Fig.~\ref{figco13}) representing the analytical model developed in Sec.~\ref{an}.

We then compute the C$^0$ mass that is needed to explain that CO mass on the bottom row of Fig.~\ref{figco13} (the red and purple lines are for $2.7 \times 10^{-3}$ and $1.2 \times 10^{-2}$ M$_\oplus$, respectively, corresponding to the range of plausible masses derived from observations). 
We find that by overplotting the best-fit contours for CO mass from the top plot on the bottom one, then for 10 Myr, the C$^0$ mass needed to explain the observed CO mass is between $5 \times 10^{-3}$ and $10^{-2}$ M$_\oplus$,
which is similar to what we deduced from the [CI] observations in Sec~\ref{gasfit} but is slightly narrower. We find that the best-fit is not necessarily at steady state. If $\alpha \sim 10^{-3}$, the disc surface density is still evolving with time as can be inferred from the fact that the contours are horizontal in this range.
We note that even for larger $\alpha$, the inner and outer regions are not necessarily at steady state (even though it is not on the horizontal part anymore) because the global viscous timescale can be $\sim 10$ times longer than the local viscous timescale for which steady state is 
reached at $R_0$ only \citep{1981ARA&A..19..137P}, i.e., the gas disc can be at steady state in the parent belt but not yet in the inner/outer regions.
As noted in Sec.~\ref{gasfit}, as the outer radial profile of the gas density is very steep, it could be that the gas disc has not reached steady state and thus the global viscous timescale should be $<15$ Myr (the age of the system), which translates as $\alpha<6 \times 10^{-3}$ as already described in Sec.~\ref{an}.
Using Fig.~\ref{figco13} and fixing $\alpha=6 \times 10^{-3}$, we therefore find a maximum $\dot{M}_{\rm CO}$ of 0.3 M$_\oplus$/Myr to still produce a best-fit (at the 99.7\% uncertainty level) in the case when steady state is not yet reached.

We note that when higher resolution images become available, the present model should be extended to a 1D code that viscously evolves CO and carbon together, since that would allow to compare the predicted radial profiles to that observed.

\subsection{Cometary composition}\label{comcomp}
 
Using Eq.~1 in \citet{2017MNRAS.469..521K}, we find that the dust production rate in this belt is $\sim 1.7$ M$_\oplus$/Myr, where we assumed a ${\rm d}r/r$ of $\sim$0.85 as derived in Sec.~\ref{dustfit}, and the fiducial values 
listed in \citet{2017MNRAS.469..521K}. This in turns leads to a prediction of the CO-to-solid mass ratio of $<20\%$ using the maximum $\dot{M}_{\rm CO}$ of 0.3 M$_\oplus$/Myr that we have just derived in the previous section \citep[see Eq. 2 in][]{2017MNRAS.469..521K}.
The upper limit we derive here is thus consistent with the largest values known for Solar System comets where the CO-to-dust mass ratio is of 3-27\% \citep{2011ARA&A..49..471M}.

As the CO-to-dust mass ratio is found to be consistent with Solar System comets, we can predict the amount of water that should be released together with CO. Using that $2 \times 10^{-3}<\dot{M}_{\rm CO}<0.3$ M$_\oplus$/Myr and 
that typical H$_2$O-to-CO mass ratios span a large range of 3-160 \citep{2011ARA&A..49..471M}, this translates 
to $6 \times 10^{-3}<\dot{M}_{\rm H_2O}< 2$ M$_\oplus$/Myr (we limit the upper bound to 2M$_\oplus$/Myr because this rate cannot be greater than the dust mass loss rate estimated above).
The water photodissociation timescale is very short and will not be much affected by the lack of UV photons with $\lambda<110$ nm (see Fig.~\ref{figphd}) because lower energy photons with $110<\lambda<190$ nm can still photodissociate H$_2$O (contrary to CO). Thus, the amount of water is expected to
be small but we can derive the amount of neutral hydrogen H that should survive together with carbon and oxygen for more than the age of the system (a few viscous timescales). Assuming that the release of CO (or that the onset of the collisional cascade) started 10 Myr 
ago (i.e., after the protoplanetary disc dispersed), we find that a mass $\sim 7 \times 10^{-3}-2$ M$_\oplus$ of H could be in the system.
 
Following a similar procedure, we also find that the total oxygen mass in the system coming from CO and H$_2$O photodissociations should be in the range 0.06-20 M$_\oplus$, which thus may be the dominant species in the system\footnote{However, we note that the upper value of 20 M$_\oplus$ is large and in the upper range compared to the total amount of oxygen expected to be available at this stage \citep[e.g.][]{2011A&A...532A..85K}}. Even if water is not released together with CO, then the oxygen mass will still be very similar to the carbon mass (times 16/12) and so would be of order $10^{-2}$ M$_\oplus$.
\citet{2015ApJ...814...42M} used Herschel to look for [OI] in HD 131835 and assuming LTE
plus that the temperature is $\sim30$ K, they find a mass upper limit of $10^{-2}$ M$_\oplus$, which is similar to our estimation when oxygen only comes from CO but slightly lower than our lower mass prediction if water is released as well. However, LTE may not be a good approximation for OI (and then the oxygen upper limit could be much higher) because the critical collider density (of electrons or H) to reach 
LTE for OI is $\sim 3 \times 10^5$ cm$^{-3}$, which may be higher than predicted by our models (especially away from $R_0$). Indeed, in Sec.~\ref{lawfit},
we found that the C$^0$ number density could be a few $10^3$ cm$^{-3}$ but owing to the low ionisation fraction, the electron density (coming from carbon photoionisation) may be rather small. If water is released, the hydrogen number density could be 10-500 times higher, i.e. the exact value of the hydrogen number density is needed to determine whether LTE is reached. If LTE is not reached, it would mean that the mass upper limit from Herschel could be order of magnitude higher in non-LTE, as already shown in \citet{2016MNRAS.461..845K}. From the observations at hand, it is thus hard to put any constraints on the release of water from the planetesimals, we can only say that the OI upper limit from Herschel is consistent with a secondary origin.

\subsection{Dust/gas interactions}\label{gasdr}
Owing to the high-CO, carbon and presumably oxygen content in the gas disc around HD 131835, the gas density may be high enough to drag the dust. To check that, we work out the dimensionless stopping time  \citep{2001ApJ...557..990T}

\begin{multline}\label{gasdrag}
T_s=8 \left( \frac{M_\star}{1.7 M_\odot}\right)^{1/2}  \left( \frac{T}{30K}\right)^{-1/2}  \left( \frac{R}{100{\rm au}}\right)^{-3/2} \\  \left( \frac{\rho_s}{1500 {\rm kg/m^3}}\right)  \left( \frac{s}{10 \mu{\rm m}}\right)  \left( \frac{n_g}{10^5 {\rm cm^{-3}}}\right)^{-1},     
\end{multline}

\noindent where $\rho_s$ is the density of the solids, $s$ the grain size, $T$ the gas temperature and $n_g$ the gas density. This equation shows that in less than 10 orbital timescales at 100 au, a dust grain of about 10 microns would be entrained by the gas and migrate radially if its density is around $10^5$ cm$^{-3}$.
Such high densities are not inconceivable close to $R_0$ in the disc around HD 131835. Indeed, assuming that the CO gas disc is 0.04 M$_\oplus$ \citep{2017ApJ...849..123M} and extends over $\Delta R=70$ au, we find that the CO number density is already $\sim 5 \times 10^3$ cm$^{-3}$. In Sec.~\ref{lawfit},
we also found that the C$^0$ number density could go up to a few $10^3$ cm$^{-3}$ on top of which some oxygen and hydrogen (coming from the release of water together with CO) would increase the total number density up to $>10^4$ cm$^{-3}$.
Hence, the gas is expected to be able to exchange angular momentum with the smallest dust grains in the system, therefore potentially leading to a radial displacement of these grains from their point of origin.

The exact interplay between the effect of gas drag and collisions is complex. Collisions will prevent grains from migrating too far (due to gas drag) as they may be destroyed before reaching their new steady state position, which needs to be modelled accurately.
Here, we simply suggest that this gas drag may be able to explain the presence of narrow rings as observed with SPHERE in scattered light \citep{2017A&A...601A...7F} because small micron-sized grains will accumulate at the outer edge of the gas disc where its density plummets \citep{2001ApJ...557..990T} 
\citep[the furthest B1 ring is at $\sim$116 au,][]{2017A&A...601A...7F} and a second ring also composed of larger grains will be left at the parent belt position \citep[the outer edge of the B2 ring is $\sim$84 au,][]{2017A&A...601A...7F}.

The increased lifetime of this small dust could also result in shielded secondary discs having on average a smaller grain size that is closer to the blow-out limit \citep[i.e., around $2.5\mu$m in this system,][]{1979Icar...40....1B} than their unshielded counterparts, evidence for which is seen in \citet{2016ApJ...828...25L}.

\subsection{Are all hybrid discs actually shielded discs of secondary origin?}\label{allhyb}

\begin{table*}
  \centering
  \caption{Table describing the disc and stellar parameters of the hybrid/shielded discs with known CO masses (calculated from optically thin CO isotopic lines).}

  \begin{threeparttable}
  \label{tabco}
  \begin{tabular}{|l|c|c|c|c|c|c|c|c|}
   \toprule
   System & $R$ & $\Delta R$ & $M_{\rm CO}$ & $T^a$ & $L_{\rm IR}/L_\star$ & age & $L_\star$ \\ 
   & (au) & (au) & (M$_\oplus$) & ($K$) & & (Myr) & ($L_\odot$)   \\ 

    \midrule
    \midrule
    HD 131835 (1,3) & 90 & 80 & 0.04 & 30 & $2 \times 10^{-3}$ & 16 & 10   \\
    HD 21997 (2,3) & 70$^b$ & 30$^b$ & 0.06 & 9 & $6 \times 10^{-4}$ & 42 & 14   \\
    HD 121191 (4,5) & 100$^c$ & 40$^c$ & $3 \times 10^{-3}$ & 6 & $2 \times 10^{-3}$ & 15 & 7   \\
    HD 121617 (4) & 83 & 60 & 0.02 & 18 & $5 \times 10^{-3}$ & 16 & 14   \\
    HD 131488 (4) & 85 & 40 & 0.09 & 10 & $2 \times 10^{-3}$ & 16 & 12   \\
   \bottomrule
  \end{tabular}
\begin{tablenotes}
      \small
	\item (a) Gas temperatures are lower limits because of possible beam dilution. (b) For HD 21997, we take the planetesimal belt radius at which 70\% of the surface density is located as being the reference radius, and the disc half-width as being from the reference radius to the inner radius derived in (3). (c) The disc around HD 121191 was not resolved (or only marginally) in (4) and thus we use an SED fit (from (5) where we added the ALMA flux) to obtain the disc radius \citep[and we correct from blackbody using][]{2015MNRAS.454.3207P}. For the disc width, we arbitrarily assume the same width as for HD 131488.
      \item References: (1) this paper; (2) \citet{2013ApJ...776...77K}; (3) \citet{2013ApJ...777L..25M}; (4) \citet{2017ApJ...849..123M}; (5) \citet{2017MNRAS.469..521K}
    \end{tablenotes}
  \end{threeparttable}
\end{table*}

\begin{figure*}
   \centering
   \includegraphics[width=13cm]{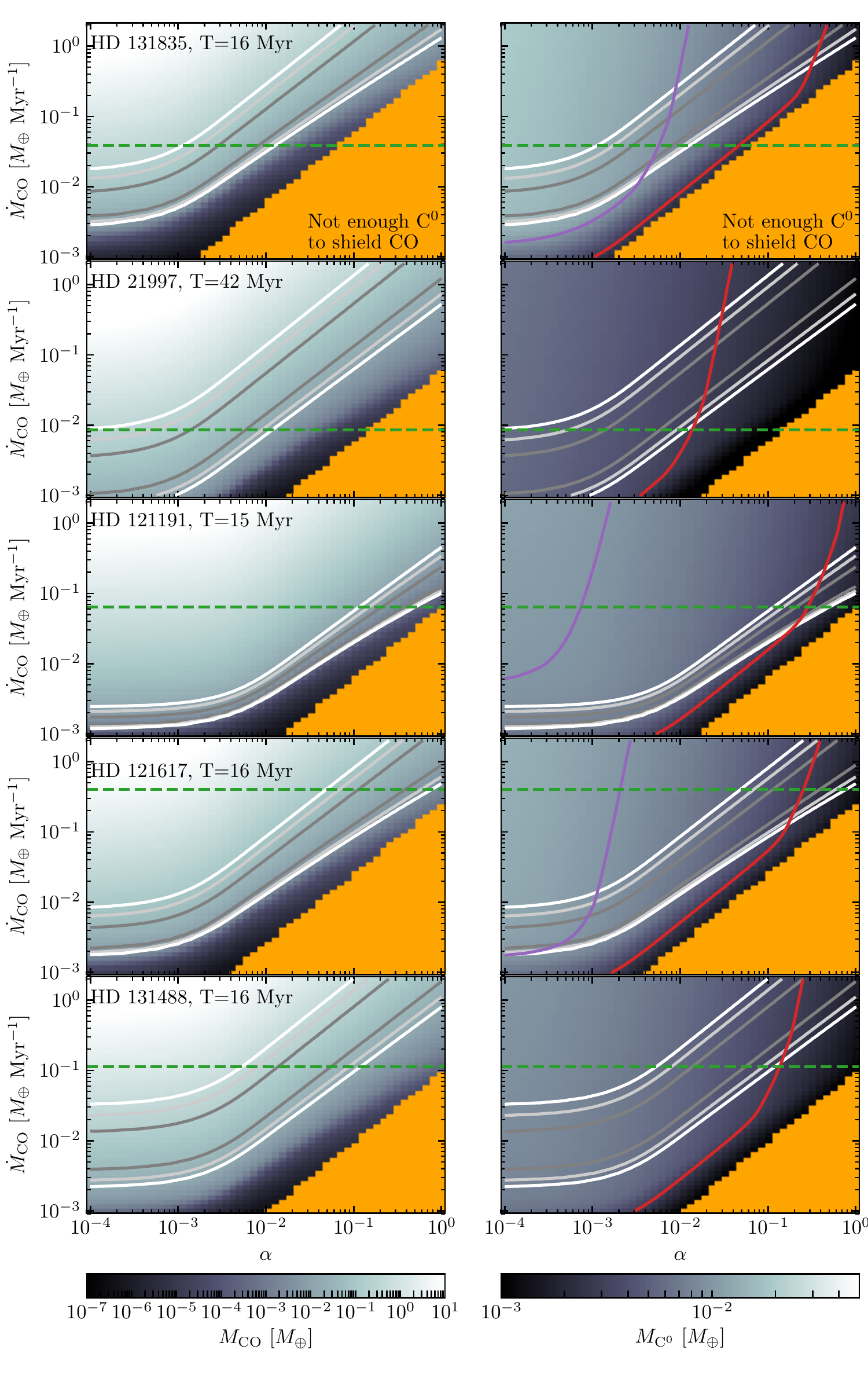}
   \caption{\label{figalpha2}  {\it Left:} Best-fit values to get the observed $M_{\rm CO}$ as observed with ALMA for HD 131835, HD 21997, HD 121191, HD 121617 and HD 131488 (see Table~\ref{tabco}), in the parameter space $\dot{M}_{\rm CO}$ VS $\alpha$ assuming that the system evolved over a 
timescale close to its age. The colourscale is showing the CO mass $M_{\rm CO}$ and the contours are 1, 3 and 5$\sigma$ from the observed value. The yellow region is where C$^0$ does not reach densities that are high enough to shield CO. 
The green dashed line shows the expected CO production rate $\dot{M}_{\rm CO}$ from the stellar and disc parameters \citep[see][]{2017ApJ...842....9M}. 
{\it Right:} Total C$^0$ mass needed to produce the observed $M_{\rm CO}$ values in different parts of the parameter space (the best fit contours for the CO mass are overplotted). The red and purple lines are for C$^0$ masses of $3 \times 10^{-3}$ and $10^{-2}$ M$_\oplus$, respectively (typical range expected in shielded discs).}
\end{figure*}

The new scenario suggested in this paper opens the possibility that all potential hybrid discs are actually shielded discs of secondary origin. So far, 9 systems (including HD 131835) have been suggested to be hybrid, 
namely HD 21997, HD 138813, 49 Ceti, HD 32297, HD 156623, HD 121191, HD 121617, and HD 131488 \citep{2013ApJ...776...77K,2016ApJ...828...25L,2017ApJ...839...86H,2016MNRAS.461.3910G,2017ApJ...849..123M}.
Out of these 9 potential hybrid members, 5 have been detected with an optically thin CO isotope line (namely HD 131835, HD 21997, HD 121191, HD 131488 and HD 121617) and thus have accurate CO measurements \citep{2017ApJ...849..123M}. 
For these, we apply the same method as described in Sec.~\ref{app} to compute the range of $\alpha$ and C$^0$ mass values that can fit the observed CO masses for each system (listed in Table~\ref{tabco}).
The results are shown in Fig.~\ref{figalpha2} in a similar way to Fig.~\ref{figco13}. 

We find that the range of $\dot{M}_{\rm CO}$ implied by the best-fitting values (contours) always overlaps with the predicted range \citep[green dashed lines from][]{2017ApJ...842....9M} derived for a given system. This reinforces the point that these systems
could be shielded discs rather than hybrid discs. In Fig.~\ref{figalpha2} (right), we also show the potential C$^0$ mass that is needed to explain the observed CO. The predicted C$^0$ masses are all above $10^{-3}$ M$_\oplus$
and would be easily detectable with ALMA, which therefore have the potential to assess the secondary origin of these gas discs.

We find that HD 131835 and HD 21997 need $\alpha$ values lower than $10^{-2}$ to explain the observations, while HD 121191, HD 121617 and HD 131488 can be explained with $10^{-2}<\alpha<1$ owing to their high CO input rate. The latter 3 systems are thus expected to be at steady state while the 
former 2 may still be evolving.

We conclude that all hybrid discs with accurate CO masses observed so far may be consistent with being shielded discs of secondary origin and that observations of [CI] with ALMA have the potential to confirm this possibility. The observed C$^0$ mass would also give insights into the most probable $\alpha$ value for each system
and thus provide information as to what is driving the viscosity in these debris disc systems \citep[e.g. whether this is the MRI, ][]{2016MNRAS.461.1614K}. We note, however, that this shielded/unshielded dichotomy seems to emerge because of differences both in $\alpha$
and $\dot{M}_{\rm CO}$. Interestingly, the $\dot{M}_{\rm CO}$ values for $\beta$ Pic and most hybrid discs are similar but the CO content observed is orders of magnitude different. 
A difference in $\alpha$ could explain these two different regimes but our model does not address the origin of the $\alpha$ values. It could be that $\beta$ Pic is actually still evolving towards that shielded stage (see the thin lines on Fig.~\ref{figcoc1alpha} that are for an evolution that has started only recently). However, it could be that $\alpha$ can be very different from system to system depending 
on the ionisation fraction or some other parameters that are yet to be understood.


\subsection{Confirming observationally whether hybrid/shielded discs are of primordial/secondary origin}\label{prim}

In this paper, we showed that massive CO gas discs may be well explained by secondary production of gas released from planetesimals in debris belts. However, can we rule out that the observed gas is primordial? 

The main difference between a primordial and secondary gas disc is that we expect the former to be dominated by H$_2$ with a typical H$_2$-to-CO ratio of $10^4$, while the latter is depleted in H$_2$ and dominated by CO, carbon, oxygen (and hydrogen, if water is released together with CO). This has several consequences which we comment on further below.

We find that it is too early to rule out a primordial origin for sure, even though, at the moment, the primordial origin is suggested based on the sole high-gas-mass content of these discs but there are no observational/theoretical proofs to back it up. Importantly, it is still to be proven how these discs could survive against photo-evaporation for $>$10Myr \citep[e.g.][]{2012MNRAS.422.1880O}?
Here, we provide some ideas for some new observations that could be used to find the real origin of the gas.

{\it Temperature}: Gas temperatures that are much lower than dust temperatures seem to be the norm for shielded/hybrid discs \citep[e.g.][]{2017ApJ...849..123M}. In a protoplanetary disc such low temperatures are not expected but it remains an open question whether in a hybrid disc scenario with low dust content such temperatures could be reached. To the authors' knowledge this has not been simulated so far. Another possibility is that the measured low excitation temperature could also be due to non-LTE effects, which would mean the actual kinetic temperature is higher \citep[e.g.][]{2015MNRAS.447.3936M}. However, as shown in Fig.~\ref{figionftdali} (right), for a gas with a secondary origin, it is possible to reach very low kinetic temperatures. We also find (using DALI) that adding sulfur in these discs can lower the temperatures even further. The exact composition of the gas disc is thus very important to set up the temperature but it goes beyond the scope of this paper to study this in detail. These low temperatures could be another hint that a secondary origin is more likely.

{\it Spatial distribution of CO VS dust and VS C$^0$}: One prediction of the model is that CO in these discs can extend further than the main planetesimal belt as it has time to spread viscously, which may explain, for instance, why the CO gas disc around HD 21997 is more extended inwards than its dust belt \citep{2013ApJ...776...77K}. Moreover, we also
predict that C$^0$ should be colocated with CO in the inner region but not in the outer region. Indeed, CO will evolve together with C$^0$ only in the regions where the C$^0$ density is high enough to shield CO beyond which the CO density would fall drastically, meaning that CO and C$^0$ would not be colocated in the outer region. In a protoplanetary disc, we expect an onion structure, with CO in the midplane and C$^0$ above it and thus CO and C$^0$ are expected to be colocated radially (though not vertically). The main observational difference between the primordial VS secondary profile is that of the steep drop of CO predicted for the secondary origin (while C$^0$ can extend further out) and thus we expect this non-colocation of CO and C$^0$ only in the secondary scenario. We note that this will happen in regions where the carbon density is smaller and deep observations may be needed to actually see the non-colocation of CO and CI. Resolving together CO and C$^0$ with ALMA would thus enable to test for consistency with a model in which the gas is secondary. 

{\it Scaleheight}: We also note that from a high-resolution ALMA image of the system, one could measure the scaleheight of the gas disc. This scaleheight depends only on the mean molecular weight of the gas (as soon as we know the gas disc temperature) and we could thus directly show that the gas is secondary/primordial from such an observation. This has already been tried for 49 Ceti where \citet{2017ApJ...839...86H} find some evidence that the scaleheight is low in this gas-rich system (hence favouring a high molecular weight and thus a secondary origin). Note that for this to work, the temperature should be known rather accurately at a specific radius where the scaleheight is measured. We could use the same method as described in Sec.~\ref{gasfit} to get the temperature using data with a higher resolution or multitransitions of optically thin species such as C$^{18}$O \citep[e.g.][]{2017MNRAS.464.1415M}.

{\it Spatial distribution of mm-dust}: From such a high-resolution ALMA image, we could also study in more detail the spatial distribution of the millimetre dust. In Sec.~\ref{gasdr}, we showed that for a secondary origin, grains of tens of microns or smaller could migrate due to the gas. However, our Eq.~\ref{gasdrag} can also be used for a primordial origin for which the gas density would go up by orders of magnitude. Assuming an H$_2$-to-CO ratio of $>10^3$, we would have a gas density that can reach $\gtrsim 10^7$ (as we had found that the CO gas density from observations is $\sim 5 \times 10^3$ cm$^{-3}$), at which point millimetre grains would start migrating as well, which could form an enhancement of dust at a specific location in the disc \citep{2001ApJ...557..990T}.

All of this together would help to confirm the origin of the gas in these systems.

\section{Summary/Conclusion}
The most major outcome of this paper is that we present  a new way to explain the so-called hybrid discs ($>$10 Myr old systems, which contain protoplanetary disc levels of CO gas) with a second generation model. In this secondary approach, molecular gas is released
from planetesimals which then photodissociates to create an atomic gas disc that spreads viscously. If the molecular gas input rate is high enough and/or that the produced atomic gas can accumulate for long viscous timescales (i.e., small $\alpha$ values), we find that the neutral carbon mass that can accumulate 
is high enough to start shielding CO (see Sect.~\ref{shield}). The massive gas discs produced by this new secondary scenario are thus not hybrid discs (for which gas is supposed to be primordial) but rather {\it shielded discs} of secondary origin as we call them throughout the paper.

By fitting the new [CI] ALMA observation around HD 131835 (see Sec.~\ref{gasfit}), we find that the C$^0$ is mostly between 40-200 au with a mass that is $2.7 \times 10^{-3}<M_{{\rm C}^0}<1.2 \times 10^{-2}$ M$_\oplus$, which is high enough to shield CO. In Sec.~\ref{an}, we present a simple analytical model that integrates shielding by C$^0$ and we compute
the $\alpha$ value needed to get the CO mass of 0.04 M$_\oplus$ observed with ALMA towards HD 131835. We find that in order to shield CO at the right level, $\alpha$ should be between $[2.5-60]\times10^{-4}$, which would create $\sim 10^{-2}$ M$_\oplus$ of C$^0$ in agreement with the observations.

We then refine the analytical model (see Sec.~\ref{refined}) and include self-shielding from CO as well as a time-dependence as C$^0$ takes time to accumulate, and we also take into account that the CO photodissociation timescale can become longer than the viscous timescale for shielded discs, meaning that CO will spread viscously before it has time 
to photodissociate. This refined model confirms our previous results and that $\alpha$ should be on the order of $10^{-3}$ to get a good fit of both CO, C$^0$ and the dust (as the CO input rate is set by the amount of
dust in the system as more dust means more CO released). We also find that the C$^0$ mass required is compatible with that observed. 

Finally, in Sec.~\ref{allhyb}, we show that this new shielded disc scenario could explain all massive (previously called hybrid) gas discs as being shielded discs of secondary origin. 
We find that the C$^0$ mass in these systems should be high and detectable with ALMA, which therefore has the potential to test our model further (and distinguish it from a primordial origin model, see Sec.~\ref{prim}) and also to constrain the viscosity in these systems, thus giving access to the possibility
of studying what is the main driver of angular momentum in these discs.

From a more detailed perspective, we find that the second generation shielded discs can have CO that is not necessary colocated with the parent belt (thus explaining resolved systems such as HD 21997). We also find that the CO density in shielded secondary discs should abruptly drop when the C$^0$ density drops below a certain threshold, thus giving the opportunity to distinguish from a primordial gas disc from high-resolution images of CO and CI. Another smoking-gun evidence to distinguish between primordial and secondary would be to measure the scaleheight of these discs (see Sec.~\ref{prim}).

By using a PDR-like model, we also find that the typical low temperature of these massive CO discs can be explained in a shielded disc scenario because of the interplay between heating by carbon photoionisation and cooling through the OI and CII lines. 

For these massive gas discs, we also find that CN, N$_2$, and CH$^{+}$ will also be partially shielded by C$^0$ and are thus expected
to be the most abundant molecular species, after CO, in these shielded discs. We note that the detection of CH$^{+}$ would lead to an estimate of the amount of hydrogen in these systems, as well as potentially the amount of water that is released from planetesimals together with CO.

We find that the total gas mass in HD 131835 may be able to drag the smallest dust grains, which may potentially explain the creation of rings similar to what observed in scattered light with SPHERE in this system. Moreover, because of the increase lifetime of small dust grains due to gas drag, we expect that shielded secondary discs should have on average a smaller grain size that is closer to the blow-out limit than their unshielded counterparts, evidence for which is seen in \citet{2016ApJ...828...25L}.

\section*{Acknowledgments}
This paper is dedicated to Manon. We thank the referee for a fair and insightful report. QK thanks Rik van Lieshout for insightful discussions about theoretical aspects of the gas model.
QK thanks Simon Bruderer for his help on the DALI code.
QK and MCW acknowledge funding from STFC via the Institute of Astronomy, Cambridge Consolidated Grant.
LM acknowledges support from the Smithsonian Institution as a Submillimeter Array (SMA) Fellow.
This paper makes use of the
following ALMA data: ADS/JAO.ALMA\#2016.1.0.01253.S. 
ALMA is a partnership of ESO
(representing its member states), NSF (USA) and NINS (Japan),
together with NRC (Canada) and NSC and ASIAA (Taiwan)
and KASI (Republic of Korea), in cooperation with the Repub-
lic of Chile. The Joint ALMA Observatory is operated by ESO,
AUI/NRAO and NAOJ. 
This work has made use of data from the European Space Agency (ESA)
mission {\it Gaia} (\url{https://www.cosmos.esa.int/gaia}), processed by
the {\it Gaia} Data Processing and Analysis Consortium (DPAC,
\url{https://www.cosmos.esa.int/web/gaia/dpac/consortium}). Funding
for the DPAC has been provided by national institutions, in particular
the institutions participating in the {\it Gaia} Multilateral Agreement.

\appendix
\section{Additional corner plot}

\begin{figure*}
   \centering
   \includegraphics[width=14cm]{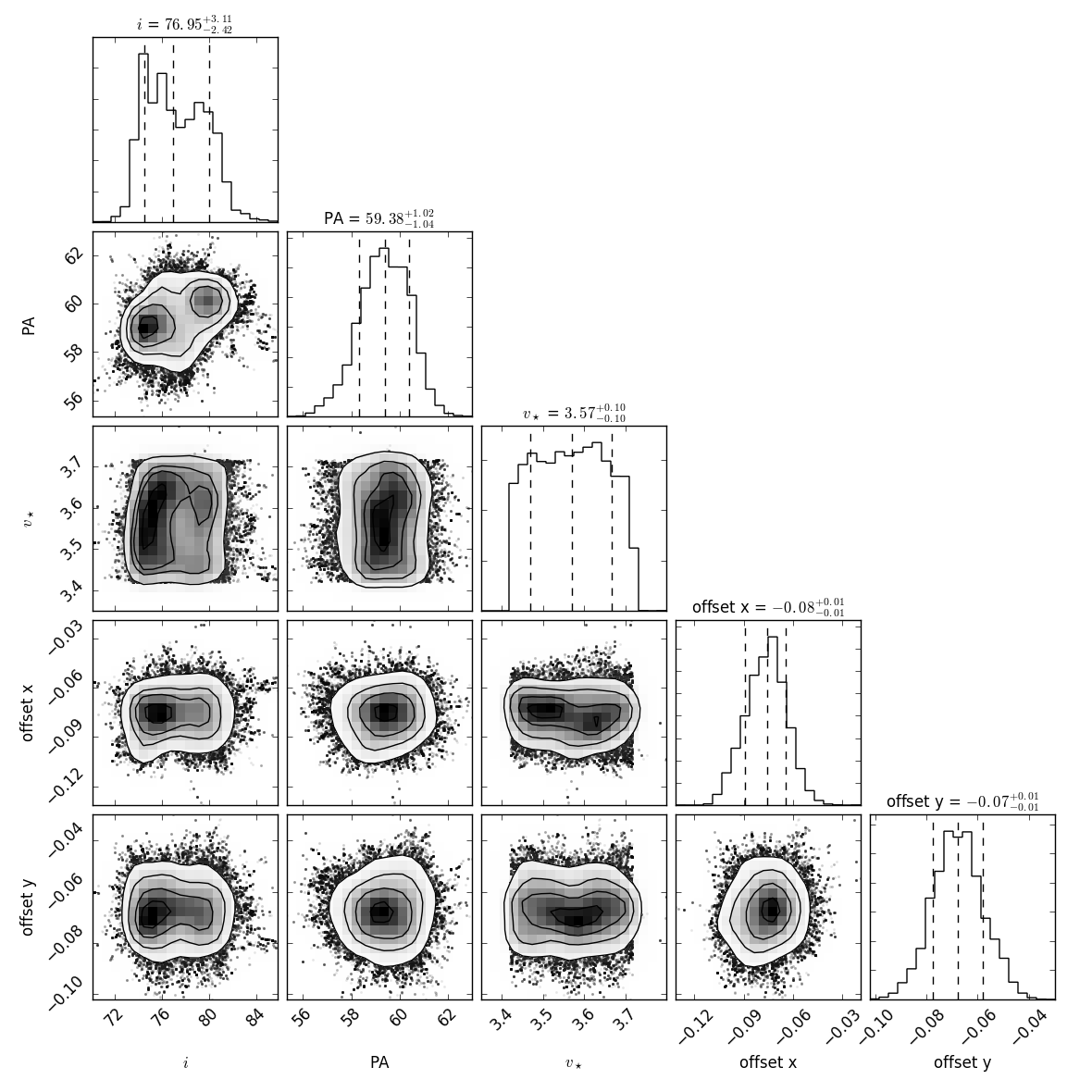}
   \caption{\label{figcorngasi} Posterior distribution for the gas disc for $i$, PA, $v_\star$, offset $x$ and offset $y$ (see Sec.~\ref{gasfit}). The marginalised distributions are presented in the diagonal. The vertical dashed lines represent the 16$^{\rm th}$, 50$^{\rm th}$ and 84$^{\rm th}$ percentiles.}
\end{figure*}




\label{lastpage}

\end{document}